# Review of Nonlinear Methods and Modelling

Frank G. Borg, Jyväskylä University, Chydenius Institute,
PB 567, FIN-67701 Kokkola (email: borgbros@netti.fi)

# Table of Contents

## Introduction

The general objective of time time series analysis is to obtain useful information from a signal regarding the state of the *source* of the signal.

Recorded biomedical signals are typically sequences (time series) of readings (measurements) $x_n$ at times $t_n$, $n = 1, 2, ..., N$ (the number $N$ of data points is commonly in the range of $10^2$ to $10^6$). The analysis of such signals is concerned with two main interconnected problems. First we have the task of *denoising* the signal; i.e., to separate (or extract) the relevant biomedical component from the noise, or from the contamination by other "disturbing" biomedical components (e.g. the eye blink from the brain signals in EEG) or the signal conditioning system. Once we have a denoised signal we may have to deal with the second task: *interpreting* the signal, which is, to answer the question: what is the meaning of the signal? One important point to emphasize here (see also Ruelle, 1994) is that the effects of the "massage" and filtering by the signal conditioning system may severly complicate the analysis of the data because it is modified by the dynamics of the filter.

In analysing signals we have two basic approaches: the dynamic approach and the statistical approach. The dynamic approaches are outgrowths of the attempts to relate the signals to the bio-physico-mathematical models of the signal source. Statistical methods are typically "ignorant" about the "mechanism" of the source and frame their task as that of finding an optimised *algorithm* (such as the ones based on ARMA-processes, see Hänsler, 1997; Moschytz & Hofbauer, 2000) for predicting new data from old data (e.g. for *control* purposes), or to compute parameters for classification. This algorithm may have no obvious relation to what may be the dynamic details of the source. Thus, the system is regarded more or less as a "black box". An interesting class of methods is based on *neural networks* (Haykin 1999) which have especially been used for *classification purposes* and *pattern recognition*, and again, not (primarily) for providing a realistic model of the system at hand.

With the advent of nonlinear times series analysis (Abarbanel 1995; Diks 1999; Kantz & Schreiber 2000; Schreiber 1998) there has been a convergence between the statistical approaches and the dynamic approaches (Tong 1990, 1995). Chaotic dynamics has shifted the emphasis from trying to determine the exact path of a dynamic system to calculating *averages* for the paths of the system. Also it has been realised, that the dynamic methods of analysis may be useful even if it cannot be directly ascertained that the system satisfies the criteria for a low dimensional deterministic system. It has been claimed (Rapp et al. 1999; Jones 2001) that a *new paradigm* has emerged within dynamic system theory - that of *algorithmic modelling*. In algorithmic modelling the starting point is the time series *per se*, not a physico-mathematical model of the system ("newtonian paradigm"). Thus, one uses the time series as the *primary data* and tries to construct an algorithmic model for this data, relevant for the purposes at hand, using methods that have been developed in the theory of dynamic systems. There are at least two theoretical justifications for this approach. First, the nonlinear dynamic theory (originating with the qualitative theory of J H Poincaré (1854 - 1912) has shown that relevant information (useful e.g. for classification purposes, not necessarily for predicting) can be obtained about a dynamic system from a time series generated by this system even if we have limited previous knowledge of the specifics of the system. Secondly, even a very complicated system (such as the heart) may operate in regimes where most of the *degrees of freedom* are "frozen" and where the system therefore behaves as a low dimensional deterministic system. This point has been forcefully demonstrated by Haken (Haken 1983, 2000) and his "school" in *synergetics*. Naturally, traditional modelling is still im-portant (Rieke et al. 1999; Keener & Sneyd 1998; Higgs & Herzog (Eds.) 1995; Koch 1999; Quarteroni 2001, 2000; Weiss 1996; Daune 1999). With regard to time series analysis such models can inform us what sort of transitions, patterns, etc., to look for in the signal, and how to effectively extract these features from the time series.

The study of nonlinear dynamic systems has also suggested the existence of so called dynamic *diseases* (Mackey & Glass 1977; Milton 2000); that is, cases where the normal functioning of a physiological system is disrupted, not primarily because of an organic failure, but because the dynamics has been shifted into a critical regime (the reason for this shift could be an improperly working control dynamics). This is compared with how changes of parameters in nonlinear dynamic system may radically alter the behaviour of the system (bifurcation, transition to chaos, etc.). Ventricular fibrillation (VF) and epileptic seizure have been proposed to be instances of dynamic diseases. (In total there are at the present some 30 - 40 types of diseases that have been tentatively identified as belonging to the class of dynamic diseases). A general observation is that healthy organism tend to have chaotic dynamics, whereas disease is related to a more regular state. As e.g. Leon Glass and Joseph Zbilut have pointed out in the context of heart studies that a chaotic dyanmics may be better prepared to swiftly respond to changing conditions.

Thus, chaos, or random fluctuations, may not necessarily be detrimental to the normal functioning of an organism. Indeed, the concept of *stochastic resonance* (Gammaitoni 1998; Bulsara & Gammaitoni 1996), originally introduced in physics by R Benzi and coworkers in 1981, has also been found to be of relevance for understanding physiological systems (Moss, 2000). Stochastic resonance works as an *amplification mechanism* which uses the *random fluctuations* of the environment in order to change the state of the system. A simple example is a particle in a potential well which, without the input of some small amount of random energy, would never be able to get out of it. A biological example would be a subthreshold neural signal which may, with the aid of a random fluctuation, exceed the threshold and thus trigger an action potential – a basic mechanism of amplification which is thought to play an important role in enhancing the sensitivity of biological signal processing systems. The somewhat conterintuitive result is that adding noise to a signal detecting system may enhance its sensitivity - a fact that probably has biological implications.

There is another similar aspect in which chaotic dynamics may enhance the vigilance of living organisms. Thus, instead of an inert stable equilibrium state, a chaotic system may have a *strange attractor* (Ruelle, 1989) such that the system visits continuously all parts of the attractor (this is connected with a so called *ergodic* property). Indeed, this is believed to be an important feature of the brain dynamics. In a sense this property of constantly tracing out its "territory" helps the system to avoid getting trapped into a dead end as suggested above. The "ergodic" property of chaotic system has an interesting consequence from the control theoretical perspective. If one wants to stabilise a certain orbit (or a fixed point) of the system, one has only to wait long enough till the system visits a neighbourhood of the desired orbit and then activate a control mechanism (e.g. by adjusting some of the parameters controlling the dynamics) which will keep the system close to this orbit. This scheme is realised by the so called OGY-control method (Ott, Grebogi & Yorke 1990). (The progress towards the desirable control point/orbit can be accelerated by the *method of tracing*.)

The central idea of algorithmic modelling is the construction of a phase space, from time series alone, using the *delay embedding* method (Packard at al. 1980) originally proposed by David Ruelle. A scalar time series $x_n$, which we suppose is generated by a (bounded) deterministic dynamics, is mapped on vectors

$$z_k = (x_k, x_{k+d}, \ldots, x_{k+(m-1)d})$$

where $d$ is the delay, and $m$ the embedding dimension. The vectors z trace out a set in the $m$-dimensional space $R^m$. If $m$ is large enough this map will generically unfold the a "diffeomorphic" copy of the original phase space (or rather the subspace where the system dwells). That means that diffeomorphic and topological *invariants* computed for the reconstructed phase space will be the same as the ones we would have computed in the original phase space had it been available to us. Thus, we may get access to characteristics of the original dynamics without any previous knowledge of the system.

A lot of research has been carried out since the 1980's, when the embedding reconstruction of phase space was first proposed, with regard to the practical applicability of the method as a basis for time series analysis. Generally, good results have been restricted to careful prepared low dimensional systems (effective dimensions in the range of $2 - 3$). In these cases there are reliable and robust methods for calculating *dimensions* (of the attractors), *Lyapunov exponents*, *entropies* and so on, as well as robust and effective algorithms for denoising and (short time) prediction (for algorithms and implementations see Hegger, Kantz & Schreiber 1998, and Kantz & Schreiber 1997). In the physiological realm claims of promising results have been reported about applications of the nonlinear time series analysis to the study of brain signals (EEG), especially in predicting the onset of epileptic seizure (Sakellares et al. 1986). In another application Schreiber and Kaplan (Schreiber & Kaplan 1996) has successfully employed nonlinear methods for separating mother and foetus ECG-signals. Measures based on nonlinear dynamics ("point correlation dimension" P2Di) are also reported to be able to discriminate, using heart beat data, patients with high risk for ventricular fibrillation (VF) among cardiac patients. The P2Di advocated by Skinner (see Skinner et al. 1998) is claimed to work for nonstationary signals. Still, *nonstationarity* is a stumbling block for most time series method used today. Another crucial problem for nonlinear time series analysis are the apparent high dimensional cases (Hegger, Kantz & Olbrich 1998) - how to distinguish a deterministic high dimensional signal from noise, and how to get reliable measures in high dimensional cases without necessarily an exponential increase in the demand on the size of data?

For classification purposes (diagnosis) the absolute meaning of the parameters are seldom of importance, only the relative orders count. Thus it might be of greater interest to track the *change* of, say, the "dimension" of a signal rather than its absolute value (which may be difficult to interpret). The fact that sometimes the order of the data, or whether it is above or below a threshold, is more important than the absolute values of the data leads to a great simplification in the form of *symbolic dynamics* (Bai-Lin, 1989). As an example, one may mark the instant when a time series crosses a threshold by 1, and the other instances by 0, thus replacing the original time series with one consisting only of zeroes and ones. Such a procedure is e.g. justified when one is interested only in the frequency of the changes for instance (typical applications are breathing patterns, heart rate variability, and the study of the synchronisation of signals such as heart rate and breathing). Generally symbolic dynamics is based on *coarse graining*; that is, dividing the phase space into a finite number of cells #1, ..., #$n$ and then recording only in which cell the dynamic system presides at a given moment. This method naturally effects considerable compactification of the data and also alleviates the sensitivity to noise.

*Determinism* means that points close to each other in phase space at a time $t$ will probably be close to each other at a time $t + \Delta t$. This property is used in the *Recurrence Quantification Analysis* (RQA) of time series (Eckmann et al., 1987). Basically one calculates the difference $|z_n - z_m|$ (using the delay embedding) which constitutes the *recurrence matrix* $R_{n,m}$. Deterministic signals will show regular patterns in the recurrence matrix. Besides offering new visual tools (recurrence plots) a number of new parameters can be defined based on RQA – one of the most important ones is the %DET-parameter ("percent determinism"). %DET is about zero for a random signal and close to 100 for a deterministic signal. RQA has been applied to sEMG, heart rate variability, breathing patterns, EEG, etc. One study (Filigoi & Felici, 1999) reports e.g. that "RQA may be proposed as a valid tool for the study of sEMG (...) on of the parameters extracted with the RQA technique, namely %DET, was shown to be a good and reliable measurement of the intrinsic structure of the signal itself". Ikegawa S et al. (2000) report a favourable comaprison between RQA-analysis and FFT-methods in assessing muscle fatigue from lumbar EMG.

In an earlier study (Nieminen & Takala 1996) applied nonlinear time series analysis to measured (by needle) myoelecric signals (MES) and compared the result with classical ARMA models. The study obtained results that "support the use of the theory of nonlinear dynamics in the modelling of the myoelectric signal". Using nonlinear prediction test (Kennel & Isabelle 1992) the MES was found to possess "a structure statistically distinguishable from high-dimensional linearly correlated noise". The embedding dimension estimated on the basis of the nonlinear predictability was in the range of 5 to 7. It is also claimed that the dimension (including the "correlation dimension") showed a significant decrease when the muscle fatigued (isometric test). The authors advocate the theory of nonlinear dynamics for "the analysis and modelling of the MES". Of course, already establishing a new significant parameter based on these methods would be an important step, irrespective of whether an explicit model can be devised or not.

Chaotic phenomena, irregular shapes etc., have often been studied by methods from fractal analysis. A recent fractal method is the *detrended fluctuation analysis* (DFA) introduced by Peng et al. (1994). It has now become one of the basic methods for analysing heart beat sequences (Peng et al. 2000; Mäkikallio 1998) and long range correlations in general. The DFA-method divides the integrated time series in blocks of size $n$ for increasing $n$. For every block one calculates the squared difference between the integrated time series and its best linear fit within the block (this involves the "detrending"). The root square of the sums of these squared differences for block size $n$ defines the "fluctuation" $F[n]$ which is often found to vary as a power of the block size $n$, $F[n] = n^\alpha$ . The exponent $\alpha$ of this power relation defines a pertinent parameter called the *self-similarity parameter* (when applicable it is related to the so-called *Hurst exponent H* by $\alpha = H + 1$). This method can be applied to non stationary series as it is insensitive to slow trends in the data. In one study of brain signals (EEG) (Hwa & Ferree 2001) they identified two scaling regions in terms of $\alpha$ and speculated that this signifies distinct dynamic processes at work.

A principal motivation behind algorithmic modelling is the fact that we may expect to encounter more and more problems where traditional modelling in terms of explicit functional relations are not available, or too complicated to implement as such. Algorithms therefore become an even more important tool, not just for handling the numerics of traditional newtonian models, but also for the very formulation of the models. The development of modern computers and electronic data acquisition systems has also made the use of the algorithmic approach feasible as well as imperative. Yet in the final end the newtonian models are also necessary if we want to understand what is going on. Algorithmic models may work as a bridge between newtonian models and experimental data. In the future we may expect a closer relation between newtonian models, algorithmic models, statistical methods, and time-frequency methods in time series analysis. For instance, we might expect an improved understanding of the molecular dynamics on the cellular level, which in turn may lead to better models for the cell interactions and the signals generated in these processes. Also, in order to be able to design optimal electrodes we need to know how the signals are generated and transmitted through the biological tissue and how they finally interact with the electrode.

# 1. Basic theoretical concepts

## 1.1 Dynamic systems

### 1.1.1 Continuous systems

Generally a continuous dynamic system consists of a manifold $M$ (phase space) and a map (evolution, flow)

$$F : M \times R \times R \to M$$

such that if the system is in the state $x_0 \in M$ at $t = t_0$ it will have moved to the point x = $F$ $(x_0, t, t_0)$ at time $t \in R$. This is commonly written also as

(1) $$x = F_{t, t_0}(x_0)$$

The Eq (1) epitomises the concept of *determinism*: If the state $x$ of the system is given at a time $t = t_0$ it is determined at any other time $t$ by the map (1). Determinism implies that evolving from time $t_0$ to $t_1$, and then from $t_1$ to $t_2$, is the same as evolving from $t_0$ to $t_2$:

(2) $$F_{t_2, t_0} = F_{t_2, t_1} \circ F_{t_1, t_0}$$

Eq (2) is called the "Chapman-Kolmogorov law". If the evolution in Eq (1) only depends on the *time difference $t_1$ - $t_0$* we have a so called *autonomous system* and instead of (1) we may write

$$x(t) = F_t(x_0)$$

which gives the state of the system at any time $t + t_0$ when it was in the state $x_0$ at (an arbitrary) time $t_0$. Eq (2) has for an autonomous system the simple form

(3) $$F_{s+t} = F_s \circ F_t$$

A simple 2-dimensional example is the rotation in the plane $M = R^2$ is given by

$$F_t(x) = \begin{pmatrix} \cos(t) & \sin(t) \\ -\sin(t) & \cos(t) \end{pmatrix} \cdot \begin{pmatrix} x_1 \\ x_2 \end{pmatrix}$$

which rotates the vector

$$x = \begin{pmatrix} x_1 \\ x_2 \end{pmatrix}$$

by $t$ radians clockwise. Typically we do not have an explicit expression for the map $F$ but at best a *differential equation* for the evolution of the system:

(4)    $$\frac{dx}{dt} = f(x, t)$$

For autonomous systems the time dependence in the RHS of Eq (4) is dropped. Since the time of Isaac Newton (1642 - 1727) such equations (4) have been the focus of theoretical research in physics (for a modern introduction to Classical Dynamics see e.g. José & Saletan 1998). Only a small ("integrable") subset of the problems in science leading to equations of motion (4) can be solved explicitly. Usually one has to further simplify the equations (by introducing approximations) or/and solve the equations numerically. This is especially true of *nonlinear* problems where the function $f$ in (4) is a nonlinear function in $x$. A classical one-dimensional simple example of a nonlinear problem which cannot be explicitly solved was given by J Liouville (1809 - 1882) (this is an example of a *Riccati differential equation* - see e.g. Simmons (1972), especially problem nr. 8 on p. 248):

(5)    $$\frac{dx}{dt} = t - x^2$$

The historically important 3-dimensional example of a dynamic system which was studied by E Lorenz in 1963 (Lorenz 1963) (as a simplified model of atmospheric dynamics) is given by

(6)
$$\frac{dx}{dt} = \sigma \cdot (y - x)$$
$$\frac{dy}{dt} = r \cdot x - y - x \cdot z$$
$$\frac{dz}{dt} = -b \cdot z + x \cdot y$$

Because of the nonlinear second order terms on the RHS of Eq (1.1.1-6) the equations can generally only be solved numerically.

### 1.1.2 Discrete systems

Discrete dynamic systems arise naturally from continuous systems when we try to solve the Eq (1.1.1-4) numerically. This usually implies a discretization of time: $t_n = n\ \Delta t$, where $\Delta t$ is the time step used in numerically integrating Eq (1.1.1-4). The simplest numerical method of integrating Eq (1.1.1-4) is the *Euler method*:

(1)
$$x_{n+1} = x_n + f(x_n, t_n) \cdot \Delta t$$
$$t_{n+1} = t_n + \Delta t$$

The smaller the time step $\Delta t$ the better, in general, will $x_n$ approximate the true value of $x$ ($n\Delta t$). Whereas the Euler method is sufficient for integrating numerical data (when $f$ is given by empirical data), there are a great number of refinements of the procedure, one of the most frequently used being the *Runge-Kutta method* of 4[th] order (Abramowitz & Steugun 1965; Harris & Stocker 1998), which can be applied when $f$ is a given analytic function.

The general definition of a discrete dynamic system comprises a set $M$ and a map ($N$ is the set of natural integers)

(2)          $F : M \times N \rightarrow M$

Given a point $x_0 \in M$ we obtain a sequence $x_n$ (an orbit) by

(3)          $x_{n+1} = F(x_n, n)$

Thus "time" is given as a discrete set of points 0, 1, 2, ... and $F$ describes the evolution for one time step. A much studied discrete nonlinear one-dimensional system is the *logistic map*

(4)          $x_{n+1} = \lambda \cdot x_n \cdot (1 - x_n)$

For the special parameter value $\lambda = 4$ the map (1.1.2-4) is termed the *Ulam map* (after S Ulam (1909 - 1984)) which exhibits fully developed *chaos*. The figure below shows the time series of the Ulam map with the initial value $x_0 = 0.3$. The time series looks random, yet it is a deterministic series produced by the simple map (1.1.2-4). From Eq (1.1.2-4) we also oberve that the evolution is *irreversible*; from $x_{n+1}$ we cannot uniquely determine the previous value $x_n$.

### 1.1.3 Chaos and attractors

Consider the simple linear one-dimensional system ($\lambda < 0$)

(1)          $\dfrac{dx}{dt} = \lambda \cdot x$

which has the solution $x(t) = F_t(x_0) = x_0 \cdot e^{\lambda \cdot t}$ . Apparently, the set of points $\{F_t(x)\}$ will approach the *equilibrium point* $x = 0$ as $t \rightarrow \infty$ for negative $\lambda$. This is an example of an *attractor* which in this case consists of a single point ($x = 0$). In a more general situation the attractor $\{F_t(x): t \rightarrow \infty\}$ may be a submanifold $S$ of the phase space $M$. Then, for a wide range of initial points $x_0$ the orbit $F_t(x_0)$ will approach $S \subset M$ as $t \rightarrow \infty$. The figure below shows the attractor of the Lorenz system for the standard parameter values ($\sigma = 10$, $r = 28$, $b = 8/3$).

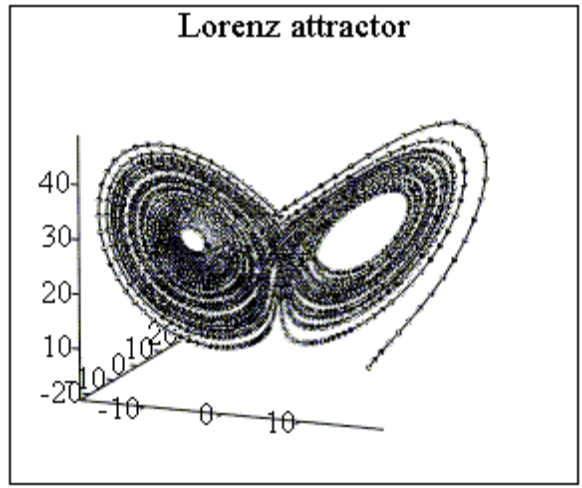

From the figure it appears as if the attractor would be a complicated 2-dimensional sub-manifold. Indeed, attractors may be complicated sets, called *fractals,* as they are associated with "fractal dimensions". When studying dynamic systems the attractors may be of intrinsic interest as their structure may give important information about the behaviour of the system. By numerically investigating Eq (1.1.1-6) Lorenz discovered that the evolution of the system had a very sensitive dependence on the initial conditions. Indeed, two orbits, $x(t)$ and $y(t)$, initially close at a time $t = 0$ tend to diverge at an exponential rate with time ($\lambda > 0$):

(2) $$|x(t) - y(t)| \approx |x(0) - y(0)| \cdot e^{\lambda \cdot t}$$

This is a defining property of *chaotic dynamics.* Of course, for a bounded system the difference (2) cannot grow beyond the size of the system. Thus (2) may be approximately valid only for restricted time spans. Anyway, the upshot is that *long range predictions* become virtually impossible. This fact had already been noted theoretically by the mathematician J Hadamard (1865 - 1963) in 1898 in connection with an investigation concerning the geodesics (curves of shortest lengths connecting two points) on manifolds of negative curvature which also lead to diverging paths. For a chaotic dynamics the state $x$ will asymptotically reach an attractor $S$, but for long time spans it will be impossible to predict *where* on the attractor the point will be. Such chaotic attractors are called *strange attractors.*

The Eq (2) is connected with the concept of *stability.* Suppose $y(0)$ is an *equilibrium point*, and thus $y(t) = y(0)$ remains fixed, then if $\lambda > 0$ Eq (2) means that a small deviation from the equilibrium leads to a growing excursion away from the equilibrium point; thus we have an *unstable* equilibrium. This is a characteristic property of chaotic systems.

### 1.1.4 Lyapunov exponents

In the limit for small $t > 0$ Eq (1.1.3-2) essentially gives the definition of a *Lyapunov exponent* $\lambda$ (after A M Lyapunov (1857 - 1918)). Consider two orbits of the system (1.1.1-4) starting at neighbouring points $x_0$ and $x_0 + \Delta x_0$. The difference (we consider the autonomous case here)

$$\Delta x_1 = F_t(x_0 + \Delta x_0) - F_t(x_0)$$

between the two orbits satisfies the equation (for small $\Delta x_0$)

(1) $$\Delta x_1 \approx DF_t(x_0) \cdot \Delta x_0$$

where $DF_t(x)$ is a matrix (the derivative of $F_t$ vis-a-vis $x$) which determines the "rate of expansion" $e^{\lambda t}$ (cmp. Eq (1.1.3-2)). By computing the average expansion along the orbit $F_t(x_0)$ we obtain the *Lyapunov exponents* $\lambda$. If we have a positive Lyapunov exponent for a bounded system, $\lambda > 0$, this indicates that the system is chaotic. Thus, the calculation of Lyapunov exponents is an important method for investigating the properties of a dynamic system. The Lyapunov exponents can be computed by directly measuring the separation of nearby orbits as a function of time and using Eq (1.1.3-2) as a definition of the Lyapunov exponents $\lambda$. For a continuous system of dimension $n$ we can define $n$ Lyapunov exponents. Indeed, imagine a small sphere $V$ of radius $r$ at the point $x \in M$. Then the flow $F_t$ of the dynamics will map, for small $t$, $V$ on an ellipsoid $V^*$ whose main axes, arranged in a descending order, are $r_1 \geq r_2 \geq ... \geq r_n$. Then the Lyapunov exponents at $x$ can be defined by (in the limit of $t \to \infty$)

(2) $$\lambda_i = \frac{1}{t} \cdot \log\left(\left|\frac{r_i}{r}\right|\right)$$

Systems with attractors are typically *dissipative systems*; that is, any volume in the phase space will contract during evolution. Given a volume $V$ in the phase space, the flow $F_t$ will map it (for $t > 0$) on a contracted volume. This is equivalent with the sum of the Lyapunov exponents being negative. For bounded continuous systems the Lyapunov exponent in the direction of the flow is *zero* if the trajectory contains no fixed point (deviations along the orbit are mapped on the orbit itself - no separation between two different orbits occurs). Thus, for a chaotic and a dissipative system we must have a dimension $n \geq 3$ since, besides the zero Lyapunov exponent, at least one Lyapunov exponent must be positive (to cause the separation of orbits), and at least one must be negative in order for their total sum to be negative.

From Eq (1.1.1-4) it can be shown that the rate of change of volumes $V$ due to the flow is given by (the "divergence" of $f$)

$$\frac{\dot{V}}{V} = \nabla \cdot f \equiv \sum \frac{\partial f_i}{\partial x_i}$$

which will be the same as the total sum of the Lyapunov exponents. For the Lorenz system (1.1.1-6) we therefore obtain

$$\lambda_1 + \lambda_2 + \lambda_3 = -\sigma - 1 - b$$

which indeed is negative for the standard choice of parameters ($\sigma = 10$, $r = 28$, $b = 8/3$). Lyapunov exponents are of special interest as they are independent of the co-ordinate system ("diffeomorphic invariant"). This follows because they are obtained by an averaging procdure. The invariance has important implications for the analysis of time series. Indeed, it means that the special representation chosen for the data does not affect the value of the global exponent. Of particular interest is the *maximal Lyapunov exponent* $\lambda_1$. A positive maximal Lyapunov exponents signals the presence of chaos. The maximal Lyapunov exponent is obtained as the exponent $\lambda$ in Eq (1.1.3-2) in the limit of $x(0) \to y(0)$ and $t \to \infty$:

(3) $$\lambda_1 = \lim_{t \to \infty} \lim_{|x(0)-y(0)| \to 0} \frac{1}{t} \log\left(\frac{|x(t)-y(t)|}{|x(0)-y(0)|}\right)$$

The reason why Eq (1.1.4-3) extracts the maximum Lyapunov exponent is because in this limit the factor $e^{\lambda_1 t}$ overshadows all the other factors determining the separation of the orbits. The formula in Eq (1.1.4-3) is also quite straightforward to implement numerically.

For discrete systems we can have chaos already in the one-dimensional case as the logistic map testifies. Consider the case of the Ulam map ($\lambda = 4$ in Eq (1.1.2-4) ). It has the *fixed point* (or equilibrium) $x^* = 3/4$. Let $w_n = x_n - x^*$ be the deviation from the fixed point, then we have

(4) $$w_{n+1} = -2 \cdot w_n - 4 \cdot w_n^2$$

Thus, $x^*$ is an unstable fixed point of the Ulam map, the deviation growing as the power of 2 with each time step near the fixed point.

### 1.1.5 Poincaré sections

There is an interesting way (introduced by Poincaré) of obtaining a discrete ($n$ - 1)-dimensional dynamics form an $n$-dimensional continuous dynamics. The case $n = 3$ may illustrate the general situation. Thus, suppose we have an orbit $x(t) = F_t(x)$ in $R^3$. A suitably placed (oriented) plane $P$ will intersect the orbit in a discrete set $D$ of points $x(t_k)$ called a *Poincaré section*. We may select those points at which the tangent of the orbit and the normal of the plane point to the same side of the plane (i.e. we may choose the points where the orbit intersect the plane from the "behind"). Since we have deterministic system the point $x(t_k) \in P$ will uniquely determine the point $x(t_{k+1}) \in P$; that is, we have 2-dimensional discrete dynamics (setting $x_k = x(t_k)$)

(1) $$x_{k+1} = G(x_k)$$

in the plane $P$. The Poincaré section contains the fundamental information about the orbit $x(t)$, such as the Lyapunov exponents (the "redundant" zero-Lyapunov exponent is eliminated in the Poincaré section).

### 1.1.6 Fractal dimensions

The attractors may have a complicated structure. An important parameter describing geometrical structures is the fractal dimension which generalizes the usual integer valued topological dimension. Mandelbrot (1982) has been the chief progenitor of the field of the fractals. The classical example (L F Richardson, 1963) is the problem of determining the true length of the rugged coast line such as that of of England. The shorter the length $l$ of the measuring stick the longer will the measure of the coast line turn out to be. Indeed, using maps of different scales it was empirically found that the total length $L$ varies as a power of the length scale $l$:

(1)  $\qquad L(l) \propto l^{1-D}$

The parameter $D$ is called the *fractal dimension* of the curve. (For the coast of France it was found that $D \approx 1.204$; for Norway, $D \approx 1.52$; for Portugal, $D \approx 1.115$.) Apparently, for a smooth line $D = 1$, agreeing with the usual topological dimension, because when the length scale $l$ becomes small enough $L(l)$ will no longer depend on $l$.

Another way to write Eq (1.1.6-1) is as

(2)  $\qquad L(l) \propto N(l) \cdot l$

where $N(l)$ is the minimum number of boxes with side $l$ needed to cover the curve. Comparing (1.1.6-2) and (1.1.6-1) we infer that

(3)  $\qquad N(l) \propto l^{-D}$

This argument can be generalized to other sets than curves. Thus, if for some set $S \subset R^m$, the minimum number of boxes with sides $l$ needed to cover the set scales as (1.1.6-3) in the limit of small $l$, then the *box-counting dimension*, or the *Kolmogorov capacity*, $D$ is defined as the limit

(4)  $\qquad D = \lim_{l \to 0} \dfrac{\log N(l)}{\log\left(\dfrac{1}{l}\right)}$

This dimension may be difficult to calculate in practice from measurement data. However, one can define (Hentschel & Procaccia 1983) a sequence of dimensions $D_0$, $D_1$, $D_2$, ... , among which $D_0$ is the same as the box-counting dimension $D$ above. Suppose again that $N(l)$ is the minimum number of *disjoint* boxes #1, #2, ...., #n ($n = N(l)$) with side $l$ needed to cover $S$. Let the number of points of $S$ contained in the box #$i$ be $N_i$ so that probability for a randomly chosen point in $S$ to be in #$i$ is given by

$\qquad p_i = \dfrac{N_i}{N}$

Then the *dimension spectrum* is defined by

(5)  $\qquad D_q = \lim_{l \to 0} \dfrac{1}{q-1} \cdot \dfrac{\log \sum\limits_{i=1}^{N(l)} p_i^{\,q}}{\log(l)}$

Apparently the definition (1.1.6-5) agrees with (1.1.6-4) in the case $q = 0$. It is possible to take the limit $q \to 1$ in (1.1.6-5) which results in the *information dimension*

$$D_1 = \lim_{l \to 0} \frac{\sum_{i=1}^{N(l)} p_i \log p_i}{\log(l)}$$

(6)

The name information dimension comes from its close relation to the *Shannon entropy* in the theory of information (Shannon & Weaver 1949). The so called *correlation dimension* $D_2$ is singled out because there are efficient algorithms for its computation (Grassberger & Procaccia 1983). One can show that $D_{q+1}/D_q$ as a function of the probabilities $p_i$ has a the maximum value 1 whence it follows that

(7)     $$D_0 \geq D_1 \geq D_2 \geq \cdots$$

Thus, $D_2$ provides a lower bound for $D_1$ and $D_0$. The name "correlation dimension" for $D_2$ comes form the terms $p_i \times p_i$ in the sum (1.1.6-5) when $q = 2$ which can be interpreted as the probability for two randomly drawn points from the set $S$ to belong to the same box #$i$. If points $\{F_{nt}(x): n \to \infty\}$ are uniformly distributed over the attractor $S$ we have

$$D_0 = D_1 = D_2 = \cdots$$

i.e. all the dimensions agree; in the opposite case we have a so called *multifractal*. A difference between the dimensions (1.1.6-5) measures the lumpiness of the distribution of the points in the attractor. Indeed, there is another way to define the sequence of dimensions (1.1.6-5). Let $N_x(r)$ be the number of points of the set $\{F_{nt}(x_0): n \to \infty\}$ which are inside a circle of radius $r$ around the point $x \in S$. Now one can show that the average of $N_x(r)$ over $S$, denoted by $<N_x(r)>$, scales like $r^{D_0}$ as $r \to 0$, and that in general

(8)     $$\left\langle (N_x(r))^q \right\rangle \propto r^{q \cdot D_{q-1}}$$

Power laws are ubiquitous in nature (Mandelbrot 1982; Schroeder 1991) and it is therefore also of great interest to investigate the presence of such "laws" in biological systems and time series.

### 1.1.7 Entropies

The concept of *entropy* stems from thermodynamics (C E G Clausius) and statistical physics (L Boltzmann) and was introduced into the theory of information by C E Shannon (1916 - 2001). In the context of dynamical systems information has to do with the amount of data needed to identify a point (i.e., the length of its "address label", or the number of bits needed to fix the point) in the phase space given a certain minimum size of resolution. If we have a set of points $x_n \in S$, $n = 1, 1, 2, ..., N,$ and cover $S$ with *disjoint* boxes #$i$ (with sides $l$) - defining a *partition P* of $S$ - each box #$i$ which contains $N_i$ points of the set $\{x_n\}$, then the average *information* in bits needed to identify a point in this set $\{x_n\}$ as belonging to a certain box #$i$ is given by *Shannon entropy* (where $p_i = N_i/N$ as before)

(1)
$$H = -\sum_i p_i \log(p_i)$$

A Rényi (1921 - 1970) defined a sequence of entropies $H_q$, $q = 0, 1, 2, ....$ by

(2)
$$H_q = \frac{1}{1-q} \cdot \log \sum_i p_i^q$$

If we take the limit $q \to 1$ we obtain the Shannon entropy (1.1.7-1). Entropies provide also invariant numbers describing the dynamics (and was for this purpose introduced into dynamics by A Kolmogorov), perhaps not as much used as the Lyapnuov exponents. Again, it is quite difficult to get reliable estimates of the entropies from experimental data due to the requirement on the size of data; the case $q = 2$ is the most tractable one from the computational point of view. $H_2$ gives a lower bound for $H_0$ and $H_1$ since we have $H_0 \geq H_1 \geq H_2 \ ...$ . More precisely, for $q > 1$ it can be shown that $H_{q+1}/H_q$ as a function of the probabilities $p_i$ satisfies (similar relation for the dimensions $D_q$)

$$1 - q^{-2} < \frac{H_{q+1}}{H_q} \leq 1$$

Indeed, the lower bound corresponds to the case $p_i \to 1$, e.g. for $i = 1$, and otherwise 0. The upper bound corresponds to the uniform distribution $p_1 = ... = p_n = 1/n$.

If we use a finer and finer partition the sums in (1.1.7-1) and (1.1.7-2) will diverge. However, there is a way of getting a finite limit by *subtracting* entropies. For this, let $p_{i_1,...,i_m}$ denote the joint probability, that for an arbitrary $n$, $x_n$ is in the box #$i_1$, $x_{n+1}$ in the box #$i_2$, etc, and define then, generalizing Eq (1.1.7-2)

(3)
$$H_q(m) = \frac{1}{1-q} \cdot \log \sum_i p_{i_1,...,i_m}^q$$

The *Kolmogorov-Sinai entropies* $h_q$ are then obtained as upper limits of the difference $H_q(m + 1) - H_q(m)$

(4)
$$h_q = \sup_P \lim_{m \to \infty} H_q(m+1) - H_q(m)$$

(the supremum limit is over all possible partitions $P$). The difference $H_q(m + 1) - H_q(m)$ gives the increase in information needed to locate a point of the set $\{x_n\}$ in an ($m$+1)-dimensional space instead of an $m$-dimensional space. The existence of the limit (1.1.7-4) means that this increment will "saturate" for high values of $m$. In dynamics the sequence $\{x_n\}$ will correspond to a time series $F_{nt}(x_0)$. The entropy will then be a measure of the rate of the separation of the orbits. Consider a small box $B$ and its transformation $F_t(B)$ by the flow as $t \to \infty$. The set $F_t(B)$ may become very complicated as time $t$ grows which means that we need more information (number of bits) in order to locate a point of the sequence $\{x_n\}$ in $F_t(B)$ than in $B$. This increase in the requirement of information for locating a point is reflected in the entropy. Entropy also gives a measure how far in the future it is possible to make predictions for the dynamical system.

## 1.2 Geometry from time series

### 1.2.1 Embedding

The arena of dynamics is the phase space. However, measurement data is usually a sequence of real numbers (a time series) $x_1$, $x_2$, ...., and we may have no information about the details of the dynamics. Assuming that the time series is determined by an unknown deterministic dynamics it is still possible, under quite general conditions, to reconstruct its phase space and thus make it possible to apply the theoretical methods for dynamical systems that were described in the previous sections. This is the foundation of *nonlinear time series analysis*. The reconstruction is based on a very simple idea. Typically the differential equation for a dynamical system is equivalent to an $m^{th}$ order differential equation

$$P\left(\frac{d^m x}{dt^m}, \frac{d^{m-1} x}{dt^{m-1}}, ..., x, t\right) = 0$$

Under quite general conditions there exists a solution $x(t)$ which is uniquely determined by the initial values (at $t = t_0$)

$$\left(\frac{d^{m-1} x(t_0)}{dt^{m-1}}, \frac{d^{m-2} x(t_0)}{dt^{m-2}}, ..., x(t_0)\right)$$

This means that a phase space for the system may be described in terms of the variables

$$y_1 = x$$
$$y_2 = \frac{dx}{dt}$$
$$...$$
$$y_m = \frac{d^{m-1} x}{dt^{m-1}}$$

For a discretizised system the differential are replaced by differences which means that for the phase space reconstruction we can use as variables (which was suggested by David Ruelle - see Packard et al. (1980) the time shifts of $x_k$:

(1)
$$y_1 = x_k$$
$$y_2 = x_{k+1}$$
$$...$$
$$y_m = x_{k+m-1}$$

The form (1.2.1-1) is a special case of a slightly more general form where we use a *delay* $d$ that can be different from 1. Thus, given a (scalar) time series $x_1, \ldots, x_N$, then its $m$-dimensional embedding in $R^m$ with delay $d$ is given by the vectors

$$(2) \quad \begin{aligned} z_1 &= \left( x_1, x_{1+d}, \ldots, x_{1+(m-1)\cdot d} \right) \\ z_2 &= \left( x_2, x_{2+d}, \ldots, x_{2+(m-1)\cdot d} \right) \\ &\qquad \cdots \end{aligned}$$

As an example, take the Ulam-time series which is graphed in section 1.1.2. If we map the time series into $R^2$ using $m = 2$ and $d = 1$ we obtain the figure below.

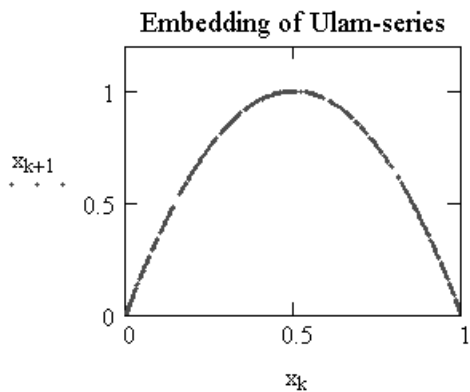

Thus the apparent random series reveals the simple structure of the dynamics when embedded in $R^2$. With real data it may not be obvious how to choose the parameters $m$ and $d$ for an optimal embedding. By chosing $m$ large enough the dynamics will indeed unfold, but an excessive large dimension $m$ will be a disadvantage when one wants to extract information from the data about dynamics since typically the requirement on the size of data grows as a power of the dimension $m$. One method for finding an optimal embedding dimension is to compute the number of *false neighbours* for $m = 1, 2, \ldots$ The idea is the following one: Two points which are close in the $m$-embedding may be so because the time series is a projection of the unfolded trajectory in a higher dimension. Two points not close in the higher dimensional space may get projected close to each other in the lower dimensional space. True neighbours though will remain neighbours when we embed the time series in a higher dimensional space, whereas the false neighbours will get "divorced". Thus, the "optimal" embedding dimension can be chosen as the dimension $m$ beyond which a further increase in the dimension does not resolve any further false neighbours. For a yet another method of determining the embedding dimension see the section 1.2.3 on nonlinear prediction.

The effect of $d$ can be illustrated by the simple sinus-wave data

$$(3) \quad x_n = \sin\left( 2 \cdot \pi \, \frac{n}{100} \right)$$

If we embed this series using $d = 1$ and $m = 2$ we obtain the graph i.e., it is almost a diagonal. However, if we use $d = 25$, which is a quarter of the period $T$ (= 100) of the wave (1.2.1-3), then we get which much more clearly shows the trajectory (circle) in the phase space than in the previous case. The point is that if we chose $d$ too small, $x_{k+d}$ will contain almost no new information compared with $x_k$. If $d$ is too large, on the other hand, $x_{k+d}$ and $x_k$ will be only related like two random numbers (if we have a chaotic dynamics). One idea for determining the delay $d$ is to calculate the *autocorrelation* for the time series,

(4)
$$\left\langle x_n x_{n+k} \right\rangle = \frac{1}{N} \cdot \sum_{i=1}^{N} x_i \cdot x_{i+k}$$

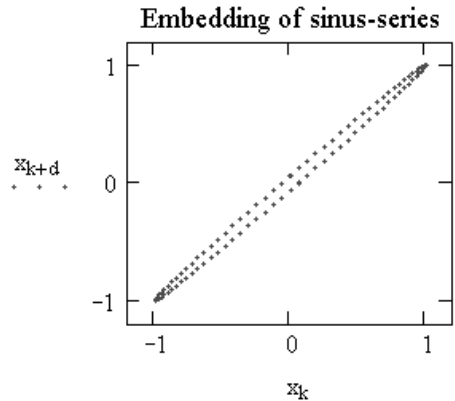

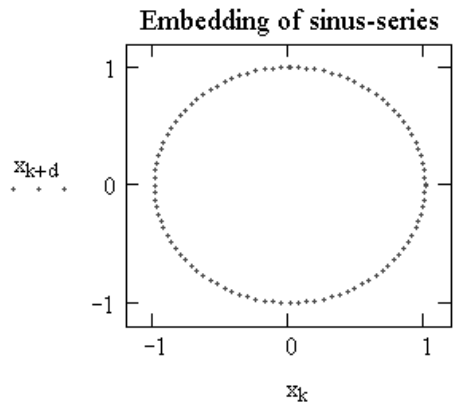

(using cyclic convention : $x_{k+N} = x_k$)

and to take $d$ as the value of $k$ where the autocorrelation has reached about $1/e$:th of its maximum value at $k = 0$. Another proposal is to use the *mutual information $I(k)$*, which is a generalization of the autocorrelation, and to take $d$ as the *first point of minimum* of the mutual information (see e.g. Abarbanel 1996). In order to define mutual information for a scalar time series $x_n$, divide its range into subintervals #$i$. Then by the histogram method calculate the probability $p_i$ for a point to be in the interval #$i$, and the joint probability $p_{i,j}$ ($d$) that if $x_k$ is in #$i$ (for any $k$) then $x_{k+d}$ is in #$j$. Finally the mutual information is defined by the expression

$$(5) \qquad I(d) = \sum_{i,j} p_{i,j}(d) \cdot \log\left(\frac{p_{i,j}(d)}{p_i \cdot p_j}\right)$$

Eq (1.2.1-5) describes the amount of information (in bits), averaged over the orbit, "shared" by the signal $x_k$ and its time shift $x_{k+d}$; that is, how much we can learn from one signal about the other. Thus, if they are independent we have $p_{i,j}(d) = p_i\, p_j$ and the mutual information is zero.

### 1.2.2 The Grassberger-Procaccia algorithm

Very useful algorithms in nonlinear time series analysis have originated with (Grassberger & Procaccia 1983). The basic algorithm consisted in computing the fraction of pairs of points in the embedded space that lie within a distance ε of each other. This defines the "correlation" $C(ε)$:

$$(1) \qquad C(\epsilon) = \frac{1}{N \cdot (N-1)} \cdot \sum_{i \neq j} \Theta\left(\epsilon - |z_i - z_j|\right)$$

where the *Heaviside function* $\Theta$ is given by

$$\Theta(x) = \begin{cases} 0 & \text{if } x < 0 \\ 1 & \text{otherwise} \end{cases}$$

From the definition of $C(ε)$ we may expect, if an attractor is present in the system, that its dependence on $ε$ reflects the fractal dimension of the the attractor. Indeed, in this case $C(ε)$ will scale as

$$(2) \qquad C(\epsilon) \propto \epsilon^{D_2}$$

where $D_2$ is the correlation dimension discussed earlier. An important improvement to the basic algorithm can be introduced by avoiding pairs of points which are too close in time, because such points tend to lie on a one-dimensional line and thus lead to an underestimation of the dimension in (1.2.2-2). This correction is effected replacing Eq (1.2.2-1) by

$$(3) \qquad C(\epsilon) = \frac{2}{(N-n_{\min}) \cdot (N-n_{\min}-1)} \cdot \sum_{i > j + n_{\min}} \Theta\left(\epsilon - |z_i - z_j|\right)$$

where all pairs closer than $n_{min}$ ("Theiler correction") in time are neglected. ($n_{min}$ is typicallly chosen to be around 500, or 10% of the total number of points.)

Corresponding to the series of entropies and dimensions discussed in 1.1.6 and 1.1.7 there is a generalization of (1.2.2-3) which will capture these parameters (for $q > 1$)

$$(4) \qquad C_q(\epsilon, m) = \frac{1}{\alpha} \cdot \sum_{i=n_{min}}^{N-n_{min}} \left( \sum_{|i-j|>n_{min}}^{N} \Theta\left(\epsilon - |z_i - z_j|\right) \right)^{(q-1)}$$

where the normalization factor is given by

$$\alpha = (N - 2 \cdot n_{min}) \cdot (N - 2 \cdot n_{min} - 1)^{(q-1)}$$

We have indicated the dependence on $m$ in (1.2.2-4) because $C_q(\epsilon, m)$ is expected to scale as

$$(5) \qquad C_q(\epsilon, m) \propto \epsilon^{(q-1) \cdot D_q} \cdot e^{-(q-1) \cdot H_q(m)}$$

As has been pointed out, the case $q = 2$ is generelly feasible. The dimensions and entropies are being estimated from (1.2.2-5) by finding *scaling regions* where

$$(6) \qquad \log\left( \frac{C_q(\epsilon, m+1)}{C_q(\epsilon, m)} \right)$$

becomes independent of $\epsilon$. The entropy can then be determined if (1.2.2-6) "saturates" with increasing $m$.

As a "rule of thumb", in order to be able to extract a dimension of value $D$ from the data the minimum size of the should be (D Ruelle)

$$N_{min} \approx 10^{\frac{D}{2}}$$

Indeed, this follows from Eq (1.2.2-1,2) if we have

$$C(\epsilon_l) \approx N^{-2}$$

at the lower end of the scaling region and

$$C(\epsilon_u) \approx 1$$

at the upper end of the scaling region, and if we require that at least $\epsilon_u / \epsilon_s \approx 10$ for the scaling region in order to be relevant. A somewhat more elaborate heuristic argument in (Hegger et al. 1998) provides the estimate

$$N_{\min} \approx e^{\frac{m \cdot h_2}{2}} \cdot \sqrt{2\ k_{\min}} \cdot \left(\frac{\epsilon_u}{\epsilon_l}\right)^{\frac{D_2}{2}}$$

Here $k_{\min}$ is the minimum number of pairs of points needed for acceptable statistics, $\epsilon_u$ and $\epsilon_l$ are the upper and lower scaling limits of the scaling region in which (1.2.2-5) is numerically valid.

### 1.2.2.1 Approximate Entropy (ApEn)

An alternative to the steps (1.2.2-1,3) in the case $q = 2$, is the basis for the definition of "approximate entropy" (ApEn) (see e.g. Yang et al. 2001) proposed in 1991 (Pincus 1991) and commonly used in the study of heart beat time series (Mäkikallio et al. 1996). In the quoted study the neighbourhood size is given in terms of a "tolerance" parameter $r$ (usually chosen to be about 0.2, i.e. 20 %, and with embedding dimension $m = 2$),

$$\epsilon = r \cdot SD$$

where $SD$ is the standard deviation of the time series. In order to compute ApEn one first calculates the quantities

$$C_r^m(i) = \frac{1}{N-m+1} \cdot \sum_{j=1}^{N-m+1} \Theta\left(r \cdot SD - \left|z_j - z_i\right|\right)$$

whose *logarithms* are averaged

$$\phi^m(r) = \frac{1}{N-m+1} \sum_{i=1}^{N-m+1} \ln\left(C_r^m(i)\right)$$

Finally the approximate entropy is defined as

(1)        $$ApEn(m,r,N) = \phi^m(r) - \phi^{m+1}(r)$$

which can be considered as an approximation of the Kolmogorov-Sinai entropy $h_2$ of Eq (1.1.7-4) (which implies a limit $m \to \infty$). For large $N$ the expression approximates the logarithm $\ln(1/p)$ of the inverse of the conditional probability $p$ that if two sets of $m$ consecutive points of the time series are close (assuming here a delay $d = 1$); that is,

$$z_k = \left( x_k, x_{k+1}, \ldots, x_{k+m-1} \right)$$
and
$$z_l = \left( x_l, x_{l+1}, \ldots, x_{l+m-1} \right)$$

satisfy $\left| z_k - z_l \right| < r \cdot SD$ then we have also $\left| x_{k+m} - x_{l+m} \right| < r \cdot SD$. Thus ApEn, like the other entropies, measures a degree of "surprise" in the data, the ability to predict the future behaviour from past patterns. The lower the value of ApEn the more regular and predictable is the data. For instance, in (Yang et al. 2001) they report a study of heart rate tracing from an infant who had an aborted sudden infant death syndrome (SIDS) and one from from an healthy infant. The more regular HRV-signal of the SIDS-infant corresponded to a lower ApEn value of 0.826 compared with 1.426 for the healthy infant. The authors emphasise the robustness of ApEn and that it "is not an approximation of the K-S entropy. It is to be considered as a family of statistics and system comparisons (...) with fixed $r$ and $m$". The ApEn is said the involve some computational advantages over the traditonal K-S entropies: less sensitive to noise, and does not require as many data points for a valid estimate (the typical size of measured biosignals may be only around 1000 points; one rule of thumb (Pincus & Goldberger 1994) is that a reliable estimate of ApEn ($m$, $r$) requires about $10^m - 20^m$ points). Also its ability to differentiate between signals with different mixes of stochastic and determinstic components is emphasized (Yang et al. 2001 p. 82 - 83).

The study (Mäkikallio et al. 1996) found that ApEn was "significantly higher in a group of postinfarction patients (1.21 ± 0.18) than in a group of healthy subjects (1.05 ± 0.11). In other study (Vikman et al. 1999) a significant decrease in the ApEn was observed before the spontaneous onset of paroxysmal atrial fibrillation (AF).

### 1.2.2.2 Sample Entropy (SampEn)

One might observe that the "Theiler correction" (see section 1.2,2) is not implemented in the definition of ApEn above so it includes a spurious contribution from "self-matching" because it counts each point $z_m$ as belonging to the neighbourhood of $z_m$. Richman and Moorman (2000) therefore suggested an alternative definition called "Sample Entropy", SampEn. Again, if we have scalar time series $x_1, \ldots x_N$, consider the $m$- and ($m+1$)-dimensional embedding vectors

$$z_k = \left( x_k, x_{k+1}, \ldots, x_{k+m-1} \right)$$
and
$$w_k = \left( x_k, x_{k+1}, \ldots, x_{k+m} \right)$$

and define

$$B_r^m(i) = \frac{1}{N-m+1} \cdot \sum_{j=1,\, j \neq i}^{N-m} \Theta \left( r \cdot SD - \left| z_j - z_i \right| \right)$$
$$B_r^m = \frac{1}{N-m} \cdot \sum_{i=1}^{N-m} B_r^m(i)$$

We then define $A_r^m$ exactly the same way using the $(m+1)$-dimensional vectors $w$ instead of $z$. Finally the estimate SampEn of the sample entropy (obtained in the limit of $N \to \infty$) is defined by

$$(1) \qquad SampEn(m, r, N) = -\ln\left(\frac{A_r^m}{B_r^m}\right)$$

Thus, in the limit $N \to \infty$ it becomes the natural logarithm of the "conditional probability that two sequences within a tolerance $r$ for $m$ points remain within $r$ of each other at the next point". Using numerical simulations Richman and Moorman (2000) demonstrated that the SampEn estimate may converge to a consistent value for considerably smaller values of $r$ and $N$ than the approximate entropy estimate. Therefore, the SampEn estimate promises to be less dependent on the data size.

### 1.2.2.3 Multiscale Entropy (MSE)

The point of departure in (Costa et al. 2002) is the general observation that a "loss of complexity .... [is] ... a generic feature of pathological dynamics", whereas many entropy algorithms "assign a higher value of entropy to certain pathological time series that are presumed to represent less complex dynamics than to time series derived from healthy function". This circumstance is attributed to the fact that these methods do not take into consideration that the biological processes may have various structures on different time scales. It is therefore suggested that we coarse-grain the data at different time scales and then compute the entropy for every scale and map its dependence on the scale. Thus, given a time series $x_1, ... x_N$, and a scale $\tau$, we define the coarse-grained time series $y^{(\tau)}$ by

$$(1) \qquad y_j^{(\tau)} = \frac{1}{\tau} \cdot \sum_{i=(j-1)\tau+1}^{j\tau} x_i$$
$$\left(1 \leq j \leq \frac{N}{\tau}\right)$$

The choice $\tau = 1$ reproduces the original times series. Next Costa et al. (2002) suggest we calculate the SampEn for the coarse-grained data for a range of scales (in the paper they calculate SampEn for $\tau = 1$ to $\tau = 20$ for data with 30000 points using parameters $m = 2$ and $r = 0.15$). Calculating SampEn as function of the scale is called the multiscale entropy (MSE) analysis. The MSE analysis is expected to reveal whether the data "contains complex structures accross multiple time scales". Thus it is found that (correlated) $1/f$-noise has a higher entropy for scales $\tau > 5$ than (uncorrelated) white noise. Studying heart beat time series it was also found that for $\tau > 6$ the healthy individuals showed higher entropy than subjects with atrial fibrillation (AF) and subjects with congestive heart failure (CHF). At $\tau = 1$ on the other hand the entropy for healthy and CHF coincided at a value much less than the entropy for AF.

### 1.2.3 Nonlinear prediction and modelling

For deterministic systems the future can in principle be predicted from past data. Deterministic chaos puts limitations on the range of meaningful prediction. Numerical predictions of weather may be of a range of one week. Our planetary system is expected to exhibit chaos on the time scale of some 5 million years (beyond that we cannot give reliable prediction about the whereabouts of the planetary orbits - for a popular review of the topic of the (in)stability of the solar system see Peterson 1995). A general implication of determinism is that points close in phase space are mapped on close points by the flow $F_t$, for $t > 0$ *sufficiently small*. In terms of an embedded time series, if $z_p$ and $z_q$ are close, then we expect also their time shifts $z_{p+1}$ and $z_{q+1}$ to be close. A simple prediction algorithm is based on this is called the *Lorenz method* - Lorenz did indeed suggest the method. As an (non practical) analogy we may think of weather forecasting. Suppose we have huge record of historical weather data. Knowing the temperature, windiness, cloudiness, etc, of the present day (and for a few days back), we may browse the past record in order to find days (or periods) with similar weather patterns in the past. We then look up how the weather in the past evolved from there on a few days ahead into the future and use that as a prediction for how the weather will be tomorrow etc.

A very simple example ($m = 1$) is the one given by the Ulam-series above. Thus given a number $0 < w_k < 1$ in a similar (unknown) series we may ask what will its next value $w_{k+1} = F(w_k)$ ($F$ is supposed to be unknown). We chose a neighbourhood size $\varepsilon$ and collect all the terms $z_n$ of a known series within the distance $\varepsilon$ of $w_k$. The average of their time shift will then be the prediction for $w_{k+1}$.

$$
(1) \qquad w_{k+1}^{\text{pred}} = \frac{1}{\#(S_k)} \cdot \sum_{z_n \in S_k} z_{n+1}
$$
$$
S_k = \{ z_n : |z_n - w_k| \le \epsilon \}
$$

Of course this procedure can be used for predicting more than one step ahead into the future. Conversely this method of prediction can be used to test the time series themselves in two ways. To beging with we divide the time series in two parts; one used for calculating the predictions, and one part used for checking the predictions so that we can calculate the error of the prediction. Now we can use this setup for determining a minimum embedding dimension $m$. We compute the error of prediction using $m = 1$, 2, .... If $m$ is less than the degrees of freedom (d.f.) of the dynamical system the error will be large. However, when we reach $m \ge$ d.f. there will be a slump in the size of error. However, if the error of prediction throughout is no less than if we had made the predictions by "tossing coins" our assumption of a determinstic process is most likely wrong. Thus, the prediction procedure also provides a method for probing whether the time series is deterministic to beging with (Kennel & Isabelle 1992).

Though the simple prediction algorithm (1.2.3-1) suffices for many purposes, more sophistcated prediction algorithms can be devised by locally approximating the dynamics using e.g. polynomial functions $p_i(z)$, $i = 1, ..., M$. Given a point $y_k$ in the embedding space, then the idea is that its time shift $y_{k+1}$ can be obtained by a function of the form

$$
(2) \qquad y_{k+1} = \sum_{i=1}^{M} c(i, k)\, p_i(y_k)
$$

Thus, Eq (1.2.3-2) constitutes a *model* of the dynamics. The coeffcients $c(i, k)$ are calculated using the standard *least square method* (LSM). We form the set $S_k$ of points of the embedded time series in an $\varepsilon$-neighbourhood of the point $y_k$ and then we determine the coefficients $c(i, k)$ such that the squared error

$$(3) \qquad \sum_{z_k \in S_k} \left( z_{k+1} - \sum_{i=1}^{M} c(i,k)\, p_i(z_k) \right)^2$$

is minimized. These coefficients will then be used in Eq (1.2.3-2) for predicting $y_{k+1}$. They constitute a *local model* of the dynamics around $y_k$. The next step $y_{k+2}$ is predicted using a local model around $y_{k+1}$, obtained in the previous step, etc. The root mean square error of this forecast, after $L$ steps, is estimated to grow as (Abarbanel 1996)

$$(4) \qquad N^{-\frac{P+1}{m_L}} \cdot e^{L\lambda_1}$$

$N$ is the number of data, $P$ is the maximum order of the approximation polynomials $p_i$, $\lambda_1$ is the maximum Lyapunov exponent, and $m_L$ is the embedding dimension (it can be less if one employs the so called *principal component analysis* to be discussed below). The sceond factor in (1.2.3-4) is exactly what we expect from the definition of the Lyapunov exponent.

An interesting application of the prediction algorithms is in the study of a possible *nonstationarity* of a time series. Suppose we divide the time series into subdivisions $J_i$ and then use the data in $J_k$ to predict the data in $J_i$ and thereafter compute the corresponding squared *mutual prediction errors* $e_{k,i}$ (see Kantz & Schreiber 1997 § 4.4). If $e_{k,i}$ changes significantly as we move away from the diagonal ($i = k$) this is a sign of a changed dynamics (nonstationarity); that is, the same dynamic rule does not apply to the whole data set.

### 1.2.4 Noise reduction

Experimental data is always contaminated by noise or other unwanted influences. A classical method of noise separation, filtering, is based on Fourier-spectrum analysis. The simplest version is the use e.g. of a *low pass filter*. It is assumed that the spectrum of the noise (e.g. high frequency vibrations) resides mainly above a certain frequency $f_u$. The signal is then denoised by dropping all the frequency components with frequency $f > f_u$. In recent decades *wavelet analysis* has become an important tool also in signal analysis complementing the Fourier-analysis. The denoising principles in wavelet analysis are similar to the ones in the Fourier case but one operates on the wavelet coefficients instead of the Fourier-coefficients of the signal. Nonlinear time series analysis provides some novel noise reduction algorithms. The simplest one is based on averaging. Let us explain it in the case $m = 3$ for a time series $x_n$. Given $x_k$ it will be replaced by an average value calculated as follows: Let $S_k$ be the set of all embedded vectors $z_i$ in an $\varepsilon$-neighbourhood of $(x_{k-1}, x_k, x_{k+1})$. Compute the local average

$$(1) \qquad \bar{z} = \frac{1}{\#(S_k)} \cdot \sum_{z \in S_k} z$$

and finally replace $x_k$ with $\bar{x}_k$ obtained from $\bar{z} = (\bar{x}_{k-1}, \bar{x}_k, \bar{x}_{k+1})$ in Eq (1.2.4-1). This method corresponds to the averaging of the vector field $f$ determining the dynamics in Eq (1.1.1-4); that is, one averages the flow over a small neighbourhood at every point in the reconstructed phase space. The hope is of course that the random fluctuations will average out. Typically the noise reduction procedure is iterated up to about eight times with decreasing ε. The effect of the noise reduction can be assessed by comparing the predicability of the cleaned data with that of the original data as described in the previous section 1.2.3.

If we think of the attractor as a surface $S$ in $R^m$, then the effect of noise is typically to spread the points above and below this surface. This suggests that noise reduction can be viewed as a geometrical problem of reconstructing the surface from its bloated version. One solution to this problem is the *local projective denoising method*. It is based on approximating the dynamics locally by linear maps. Geometrically it means that we approximate locally the orbit (attractor) by its tangent space. The noise spreads the point below and above the "true" tangent space. In the projective noise reduction method we try to determine the local tangent space and project the points in the neighbourhood, which have drifted away, back on to the tangent space. Thus, in the cleaned data the points are replaced by their local projections. If $m_0$ is the smallest integer equal to or larger than the dimension of the attractor, and the data is embedded in a space of dimension $m > m_0$, then we may suppose that the noise effects a drifting of the points evenly in the $q = m - m_0$ orthonormal directions $a^i$ of the "null space". The projection onto the tangent space is then given by

$$P(z) = z - \sum_{i=1}^{q} (z \cdot a^i) a^i$$

Again we invoke the least square method in order to determine the orthonormal vectors $a^i$ of the null space. Indeed, we determine $a^i$ such that the squares of the distances of the points in the neighbourhood from the tangent space will be a minimum; this is a generalization of the usual linear regression method of fitting a straight line to set of points and is the content of the *principal component analysis* (PCA). The point of PCA is reduce the dimension of the data space to a subspace where its main variation resides. Now, our problem is reformulated as the minimization of the expression

(2) $$\sum_{i, z \in S_k} (a^i \cdot z)^2 - \sum_i \lambda_i a^i \cdot a^i$$

with respect to the vectors $a^i$. The last term (2) is due to the constraint that the vectors $a^i$ be of unit magnitude. The minimization of (2) gives the eigenvalue equation

(3) $$C a^i = \lambda_i a^i$$

where the symmetrical positive definite $m \times m$ "covariance matrix" $C$ is defined by

(4)
$$C_{rs} = \sum_{z \in S_k} z_r z_s$$

(here $z_r$ denotes the $r$:th co-ordinate of the point $z$ in $R^m$). The vectors $a^i$ correspond to the the $q$ eigenvectors of (1.2.4-6) with the *smallest* eigenvalues $\lambda_i$. In practice the algorithm is modified by replacing $z$ above in Eqs (2-4) with its deviation from the local average

(5)
$$w = R(z - \bar{z})$$

where $\bar{z}$ is the average ("center of mass") of the embedded points in the $\varepsilon$-neighbourhood of $z_k$. $R$ is a *weighing matrix* typically chosen (if $m > 1$) as a diagonal matrix with the elements (chosen so that $\sum_i \frac{1}{R_{ii}} = 1$ , while $\delta$ is small number such as $10^{-4}$)

(6)
$$R_{ii} = \begin{cases} \dfrac{2\ \delta + m - 2}{\delta} & \text{if } i = 1 \text{ or } m \\ 2\ \delta + m - 2 & \text{otherwise} \end{cases}$$

The point of the weighing matrix is to suppress the contributions by the end co-ordinates of the embedding vectors because they accumulate the greater part of the instabilities. A point $x_k$ of the times series is part of $m$ embedded vectors

(7)
$$z_k^{(1)} = \left( x_k, x_{k+1}, \ldots, x_{k+m-1} \right)$$
$$\cdots$$
$$z_k^{(m)} = \left( x_{k-m+1}, x_{k+1}, \ldots, x_k \right)$$

and those of them where $x_k$ is in the middle of the embedding vector will be emphasized using (6) by weighting the correction to $x_k$ from the embedding vectors (7) with $1/R$ (see Eq (9)). The projective correction replaces the embedding vector $z_n$ with

(8)
$$z_n^{\text{corr}} = z_n - R^{-1} \sum_{i=1}^q a^i \left( a^i \cdot R(z_n - \bar{z}) \right)$$

and the correction to $x_k$ is calculated from the weighted mean $\bar{z}$ of the corrected versions of the embedded vectors in (7)

(9)
$$\bar{z} = \sum_{i=1}^m \frac{1}{R_{ii}} z_k^{(i)\text{corr}}$$

The nonlinear denoising algorithms described above are of the "blind" type. No knowledge is presupposed about the details of the dynamics. This is typically the case in biological applications. In physical laboratory experiments the situation may be different if one probes accurately modelled systems. Then one can use the explicit model of the dynamics in devising the denoising algorithm (see e.g. Abarbanel 1995 ch. 7). Similarly, if one has a clean sample of a signal generated by the source one can use this for a statistical denoising of its contaminated signals.

An interesing and successful application of the nonlinear projective noise reduction method is presented in (Hegger et al. 2000). They show that the method works on denoising human speech signals though such signals are neither deterministic nor stationary. However, it seems as if the speech signals were approximately deterministic and low dimensional on the phoneme level. In the example cited the authors used embedding dimensions of the order 20 - 30 and a projection to a low dimensional subspace ($q$ in the range of 3 to 7). The neighbourhood size was chosen so as to guarantee 5 to 20 neighbours in the sets $S_k$ above. The large embedding dimensions made the method quite resilient with respect to the noise.

Denoising can be used for *signal separation*. Suppose a signal is superposed by a weaker background signal. Consider first the background signal as "disturbance" and clean it away by denoising the signal. Finally, retrieve the weak signal by subtracting the cleaned signal from the original signal and denoise it. Schreiber and Kaplan (1996) have used this method together with the projective denoising method for separating mother and foetus ECG. This method works with data from only one channel, whereas e.g. the *independent component analysis* (ICA) (Hyvärinen & Oja 1999) may require several channels of recordings in order to separate signals by combining inputs from e.g. different locations.

### 1.2.5 Detrended fluctuation analysis (DFA)

Many signals of biological origin exhibit an absence of characteristic temporal and/or spatial scales (see e.g. Hastings & Sugihara 1993; Schroeder 1991). Thus it might be that a process $x(t)$ and its time-scaled version $x(at)$ have the same statistics (averages, variances etc) if we rescale the signal itself too:

(1) $$x(t) \overset{d}{=} \frac{1}{a^H} \cdot x(at)$$

The meaning of the "distributional equvivalence" Eq (1) is that[1] the probabilities satisfy the scaling relation

(1*) $$Prob(x(t) < x) = Prob\left(\frac{1}{a^H} x(at) < x\right) \quad .$$

A robust method for studying such fractal (or self-similar) behaviours of signals was introduced by Peng et al. (1994) called the *detrended fluctuation analysis* (DFA). The property (1,1*) implies that the variance scales as

---

1   For a discussion of this statistical equivalence concept see Beran (1994) p. 50. A process that satisfies
    (1) is said to be *H-self-similar* (H-ss).

(2) $$\left\langle (x(t+\Delta)-x(t))^2 \right\rangle = \sigma^2 \left| \Delta \right|^{2H}$$

(< > denotes averaging over the orbit or a great number of runs). Here $\sigma^2$ is called the variance and using the common assumption that $x(0) = 0$ we have in general for the second moment[2]

(2*) $$\left\langle x(t)^2 \right\rangle = \sigma^2 \left| t \right|^{2H}$$

In principle the parameter $H$ can be determined from empirical data by calculating the logarithm of the LHS of (2) for a series of time differences $\Delta$ and plot it against the logarithm of $\Delta$. The DFA-method modifies this procedure. First, it is not the signal $x$ itself which is used but its *integral*[3] defined by (for discrete signals)

(3) $$y_k = \sum_{i=1}^{k} x_i$$

Secondly, suppose the size of the data is $N$. Then we divide the integrated data $y$ into nonoverlapping blocks of size $n < N$. For every such block we determine its linear least-square fit, which together define a picewise linear curve $yn$. Subtracting this from $y$ we get a "detrended" signal, and the "fluctuation" at level $n$ will be defined by

(4) $$F[n] = \sqrt{\frac{1}{N} \cdot \sum_{i=1}^{N} \left( y_i - yn_i \right)^2}$$

If $x$ satisfies the scaling property (1,2) then $F[n]$ will follow a "power law"

(5) $$F[n] \propto n^{\alpha}$$

The exponent $\alpha$ in Eq (5) is termed the *self-similarity parameter*. Generally, a higher $\alpha$-value implies a more regular time series. It is related to the *Hurst parameter H* defined by Eq (2). Since integration raises scaling exponents of the signal by one, the Hurst parameter and the self-similarity parameter are related by (for fractal Brownian motion)

(6) $$\alpha = H + 1$$

Indeed, from (1) it follows for

$$y(t) = \int_0^t x(u) \, du$$

---

2    The properties $x(0) = 0$ and (2, 2*) define *fractal Brownian processes*.
3    Reduces the effect of noise.

that we have

$$y(at) = \int\limits_0^{at} x(u)\,du = a\int\limits_0^t x(au)\,du \overset{d}{=} a^{H+1}\,y(t) \quad .$$

Given a fractal Brownian process we obtain for the correlation function, using (2, 2*),

$$(7) \qquad C_t(\tau) \equiv \langle x(t+\tau)\,x(t)\rangle = \frac{\sigma^2}{2}\left(|t+\tau|^{2H} + |t|^{2H} - |\tau|^{2H}\right)$$

In the limit $\tau \to \infty$ $(\tau \gg t > 0)$ this is approximated by

$$(7^*) \qquad C(\tau) \approx \frac{\sigma^2}{2}(2H+1)\cdot\tau^{2H-1} \quad .$$

That is, we have

$$(8) \qquad C(\tau) \propto \tau^{-\gamma} \ \text{ for } \ \tau \to \infty$$

with $\gamma = 1 - 2H$. In the next section (1.2.5.1-20) it will be shown that the power spectrum $S(f)$ scales as

$$(9) \qquad S(f) \propto \frac{1}{f^\beta}$$

with the exponent $\beta$ given by[4]

$$(10) \qquad \beta = 2H + 1 = \alpha + 1 \quad .$$

There is a characteristic difference between signals with $0 < H < 0.5$ and $H > 0.5$. The former is associated with *anti-presistence* whereas we later is connected with *persistence*. Persistence means that for time intervals $\Delta t$ the changes $x(t + \Delta t) - x(t)$ and $x(t) - x(t - \Delta t)$ tend to have the same sign (changes enforce themselves), while in the case of anti-persistence they tend have opposite signs (changes provoke changes in the opposite direction). More precisely, define $\rho$-correlation $-1 \le \rho \le 1$ for a time scale $\Delta t$ by

$$(11) \qquad \rho = \frac{\langle(x(t+2\,\Delta t) - x(t+\Delta t))\cdot(x(t+\Delta t) - x(t))\rangle}{\sqrt{\langle(x(t+2\,\Delta t) - x(t+\Delta t))^2\rangle\cdot\langle(x(t+\Delta t) - x(t))^2\rangle}}$$

---

4   Observe that (8) could wrongly suggest that $\beta = 1 - \gamma = 2H$.

Now, using the definition (2) of the *Hurst exponent H* we get

$$(12) \qquad \left\langle (x(t+2\,\Delta t)-x(t))^2 \right\rangle = \text{const.}\cdot 2^{2H}|\Delta t|^2 H$$

Combining this with (11) and (2) we obtain

$$(13) \qquad \begin{aligned} 2\,(1+\rho)&=2^{2H} \\ (0 &< H < 1) \end{aligned}$$

From this we indeed see that the $\rho$-correlation (11) is negative for $H < 0.5$. Zero $\rho$-correlation corresponds to the "classical" Brownian random walk. The Hurst exponent can also be directly related to the *fractal dimension D* of the graph $(t, x(t))$ of the time series in the plane by (see e.g. Hastings & Sugihara 1993)[5]

$$(14) \qquad D = 2 - H$$

Thus the Brown time series $x(t)$ ($H = 0.5$) has the fractal dimension $D = 3/2$. Fractal time series can furthermore be characterised by their *zero-crossing* probabilities. Suppose we have $x(0) = x_0$, then what is the probability that $x(t) > x_0$ for $0 < t < T$ and $x(T) = x_0$? The probability density can be shown (Ding & Yang 1995) to be related to the fractal dimension $D$ and is inversely proportional the $D$th power in $T$:

$$(15) \qquad P(T) \propto T^{-D}$$

a relation that has also been used for numerically determining the fractal dimension $D$ and thus the Hurst exponent for long time series (Liebovitch & Yang 1997).

Numerous studies have applied DFA to heart beat interval data. In one study it was found (see Peng et al. 2000) that for a group of people with no known heart problems the self-similarity parameter $\alpha$ was about $1.0 \pm 0.1$ whereas for a heart failure group it was about $1.24 \pm 0.22$.

### 1.2.5.1 Disgression: Fourier transform and power laws

For a continuous[6] stationary *stochastic process* $x(t)$ we define the power spectrum $S(f)$ by

$$(1) \qquad \langle \hat{x}(f)\overline{\hat{x}}(v)\rangle = \delta(f-v)S(v)$$

---

5  Heuristically, when covering the curve with boxes of reduced size $\Delta x/2$ the number of boxes $N(\Delta x/2)$ must be increased on the average, due to (1, 12), by a factor of $2^{1\cdot H}$ as compared to the case of a smooth line ($H = 1$); this gives $N(\Delta x/2) = 2^{2\cdot H} N(\Delta x)$ and thus $D = 2 - H$ (by the definition (1.1.6-3) ).

6  In the discrete versions the Dirac-delta functions are replaced by Kronecker-deltas and a constant dimensional factor.

(in time domain this is equivalent to

(1*) $\qquad \langle x(s)x(t) \rangle = g(s-t)$

where $g(t)$ is the Fourier-inverse of $S(f)$.) From this we obtain for the correlation function

$$c(\tau) = \langle x(t+\tau)x(t) \rangle =$$
$$\int \int \langle \hat{x}(f)\overline{\hat{x}}(v) \rangle e^{i2\pi(f(t+\tau)-vt)} df\, dv =$$
$$\int e^{i2\pi f\tau} S(f) df$$

that is, the Fourier transform of the correlation function $C(\tau)$ is equal to the power spectrum,

(2) $\qquad \hat{c}(f) = S(f)$ .

If we take the inverse Fourier transform of a power spectrum of the variety (1.2.5-9) we arrive at an expression of the form

(3) $\qquad c(\tau) = \int\limits_0^\infty \dfrac{\cos(2\pi f\tau)}{f^\beta} df$

In the case $0 < \beta < 1$ the integral is convergent and by making the variable substitution $f\tau \rightarrow u$ it becomes

(4) $\qquad c(\tau) = \tau^{\beta-1} \int\limits_0^\infty \dfrac{\cos(2\pi u)}{u^\beta} du$

showing that $c(\tau)$ scales as $\tau^{\beta-1}$.

Spectrum power law with $\beta = 1$ is a special case. The corresponding time series is called *pink noise* (or *flicker*), or simply $1/f$-noise, and has drawn a lot of interest from researchers (see e.g. Mandelbrot 1999). In order to calculate the correlation function for the $1/f$-noise we will calculate the difference $c(\tau_2) - c(\tau_1)$ in terms of the integral

(5) $\qquad c(\tau) = \int\limits_\epsilon^\infty \dfrac{\cos(2\pi f\tau)}{f^\beta} df$

and let $\varepsilon$ go to zero. This results in (the last equality is valid for small $\varepsilon$)

(6) $$c(\tau_2) - c(\tau_1) = \int\limits_{\epsilon \tau_2}^{\epsilon \tau_1} \frac{\cos(2\pi u)}{u}\,du \simeq \ln\left(\frac{\tau_1}{\tau_2}\right)$$

Thus, for pink noise the correlation function $C(\tau)$ varies as the logarithm $\ln(1/\tau)$ (plus a "calibration constant" $C(1)$).

The above procedure of introducing a lower cut-off $\varepsilon$ for the frequency can also be interpreted in terms of the *renormalization* procedures used in physics. Another approach is based on calculating the Fourier transform of the so-called *principal value* of $1/|f|$ in terms of *distributions* (generalized functions). For a discussion see Saichev and Woyc­zyñski (1997), p. 126 – 130. For $\beta > 1$ Eq (3) defines a divergent integral and introducing a cut-off produces a somewhat arbitrary results in contrast to the $\beta = 1$ case. For $\beta$ in the range $1 < \beta < 3$ we may instead "renormalize" (3) by subtracting the infinite constant $c(0)$ from it thus obtaining a convergent integral

(7) $$c_{ren}(\tau) = \int\limits_0^\infty \frac{\cos(2\pi f \tau) - 1}{f^\beta}\,df$$

Again making the variable substitution $f\tau \rightarrow u$ we get

(8) $$c_{ren}(\tau) = \tau^{\beta-1} \cdot \int\limits_0^\infty \frac{\cos(2\pi u) - 1}{u^\beta}\,du$$

Thus the "observable correlation" function (for $1 < \beta < 3$) can be expected to be of the form

(9) $$c_{obs}(\tau) = c_{cal} - a \cdot \tau^{\beta-1}$$

where $c_{cal}$ is a calibration constant and the parameter $a$ is given by

(10) $$a = \int\limits_0^\infty \frac{1 - \cos(2\pi u)}{u^\beta}\,du = (2\pi)^{\beta-1} \cdot \int\limits_0^\infty \frac{1 - \cos(u)}{u^\beta}\,du$$

Thus if we expect a power spectrum coeffcient $\beta$ to be in the interval $1 < \beta < 3$ it could be estimated from the correlation data by fitting it to a curve of the form (9) for small time values $\tau$. (Not a very reliable method however.) For power spectrum coeffcients $\beta$ in the range $3 < \beta < 5$ similar methods suggest an observable correlation function of the form

(11)  $\qquad c_{\text{obs}}(\tau) = c_0 - c_1 \cdot \tau^2 - c_2 \cdot \tau^{\beta-1}$

which means that one cannot detect the presence of a power exponent $\beta > 3$ since it is dominated by the second order term for small $\tau$. The Fourier transform of $\tau^2$ in the generalized sense is $\quad -\delta^{(2)}(f) \quad$ (the second derivative of the Dirac delta). The Fourier transform of $|\tau|^{\beta-1}$ for non-integral $\beta > 0$ is given by

(12)  $$\frac{2\ \Gamma(\beta)\cos\left(\beta\dfrac{\pi}{2}\right)}{\left(2\ \pi\ f\right)^{\beta}}$$

where $\Gamma$ is the gamma-function defined for $\beta > 0$ by

(13)  $$\Gamma(\beta) = \int_0^\infty u^{\beta-1}\, e^{-u}\, du$$

The result (12) is derived from the expression (13) by allowing complex integration variables in (13).

The fractal Brownian process (1.2.5-2,2*) is not a stationary process since the correlation function $C(\tau)$ (1.2.5-7) depends also on $t$. Thus, we cannot directly use the relationship (2) between the power spectrum and the correlation function. However, the increment

(14)  $\qquad z(t) = x(t) - x(t-\Delta)$

is found to be stationary[7], since

(15)  $\qquad \langle z(t+\tau)z(t)\rangle = \dfrac{\sigma^2}{2}\left(|\tau-\Delta|^2 H + |\tau+\Delta|^2 H - 2\,|\tau|^2 H\right)\quad.$

From (15) we get for $\tau \gg \Delta$ the approximation

(15*)  $\qquad \langle z(t+\tau)z(t)\rangle \approx \sigma^2\,\Delta^2\,H(2H-1)\cdot\tau^{2H-2}$

Taking the Fourier transform of (15), and using the result (12), we obtain for the power spectrum of the increment $z$,

---

7   Eke et al. (2000) make a point about distinguishing between stationary fractal Gaussian noise (fGn), like the increment $z$ (14), and nonstationary fractional Brownian motion (fBm), similar to the process $x$ above. The authors emphasize that Fourier methods may be poor in distinguishing between these categories for experimental data.

(16)     $$S_z(f) = 4 \; \sigma^2 \, \Gamma(2H+1) \sin(\pi H) \sin(\pi f \Delta)^2 \, \frac{1}{(2 \, \pi \, f)^{2H+1}}$$

For "small" $f$, that is, $f < 1/\pi\Delta$, (16) implies that

(17)     $$S_z(f) \propto \frac{1}{f^{2H-1}}$$

In the same limit we may think of the increment $z$ (14) as a differential of the process $x$ whence the "power spectrum" $S(f)$ of $x$ should be related to $S_z(f)$ by $\;S(f) \approx S_z(f) f^{-2}\;$ in agreement with (1.2.5-9). Indeed, the increment $z$ can be expressed as a convolution,

(18)     $$z(t) = \int \delta_\Delta(u)\, x(t-u)\, du$$
$$\delta_\Delta(u) = \delta(u) - \delta(u-\Delta)$$

From this we get

(19)     $$\hat{z}(f) = \hat{x}(f)\left(1 - e^{-i2\,\pi f \Delta}\right)$$

which in combination with (16) "suggests"

(20)     $$S(f) = \langle |\hat{x}(f)|^2 \rangle = \sigma^2 \, \Gamma(2H+1) \sin(\pi H) \cdot \frac{1}{(2 \, \pi \, f)^{2H+1}} \quad .$$

This is interpreted as a sort of "averaged" power spectrum for the fractal Brownian motion (Flandrin 1992, 2002).

After this mathematical exercise it is to be emphasized that we may hardly expect real data to adhere to the pure models of fractal noise or motion[8]. Selfsimilarity in nature is only possible over a restricted range of scales due e.g. to the discrete constitution of matter. Therefore it is advisably to use complementary methods when e.g. trying to extract Hurst exponents from the data (see e.g. Eke et al. 2000).

---

8  "If course, true fBm are unlikely to be observed in natural phenomena, but the fBm model offers an extremely powerful idealization and a very convenient starting point for studying statistically self-similar processes, as does, for example, white noise in the case of stationary processes." (Flandrin 1998).

### 1.2.5.2 Rescaled Range Analysis (R/S)

This method for studying power laws of time series goes back to Hurst (1951). Suppose we have time series $x_i$, $i = 1, \ldots, N$ and divide it into $m$ blocks with $n = N/m$ points each. For each $1 \leq j \leq m$ define the vectors,

$$(1) \qquad \xi_j[k] = x_{(j-1)n+k}$$

and the cumulative sums

$$(2) \qquad y_j[k] = \sum_{i=1}^{n} \xi_j[i] \quad \text{for } (1 \leq k \leq n)$$

and the "accumulation"

$$(3) \qquad z_j[k] = y_j[k] - k\, \overline{\xi}_j$$

where $\overline{\xi}_j$ is the average of the vector $\xi_j$. One then defines the range $R_j(n)$ as

$$(4) \qquad R_j(n) = max_k\, z_j[k] - min_k\, z_j[k]$$

and the scale $S_j(n)$ as the standard deviation of $\xi_j$. The final range $R(n)$ for the time series and for the block-size $n$ is defined as the average of (4) over $j$. Similarly for the scale $S(n)$. It was then the observed by Hurst that the so called *rescaled adjusted range* $R(n)/S(n)$ for his data tended to scale as a power of the block size $n$,

$$(5) \qquad \frac{R(n)}{S(n)} \propto n^H$$

The relationship (5) can be argumented for if we assume a scaling property (1.2.5-1) for the data (Beran 1994 § 4.2). The R/S method may be useful for preliminary visual inspection possible long-memory trends in the data but it is not considered trustworthy for automated analysis.

### 1.2.5.3 Second order moving average

Given a time series $x[i]$, $i = 1, \ldots, N$, we may conisder the moving averages

$$(1) \qquad y_T[k] = \frac{1}{T} \sum_{i=0}^{T-1} x[k-i] \quad \text{for} \quad T < k \leq N$$

for a certain time "window" $T$. This is similar to the coarse-graining method encountered in § 1.3.3.3. Given two time windows $T_1 < T_2$, and calculating how often the curves $y_{T1}$ and $y_{T2}$ cross each other using artificial fractal data, Vandewalle and Ausloos (1998) found an empirical estimate for the density $\rho$ of the crossing points

(1) $$\rho \propto \frac{1}{T_2} \left( \Delta T \left( 1 - \Delta T \right) \right)^{H-1}$$

where

(2) $$\Delta T = \frac{T_2 - T_1}{T_2}$$

and $H$ is the Hurst exponent of the simulated data. The "empirical" formula (1), though lacking theoretical support, suggests a method for estimating the Hurst exponent from data and it seems to be comparable to the DFA-method as far as it has been tested by Vandewalle and Ausloos. Inspired by these results Alessio et al. (2002) proposed to analyse the statistics of the quantity

(3) $$\sigma_{MA}(T) = \sqrt{\frac{1}{N-T} \sum_{i=T}^{N} \left( x[i] - y_T[i] \right)^2}$$

Using simulated data they found that

(4) $$\sigma_{MA}(T) \propto T^H$$

This agrees with the theoretical result (Raymond and Bassingthwaighte 1999)

(5) $$\langle SD_T^2 \rangle \approx \frac{\sigma^2}{2} \frac{T^2 H}{(2H+1)(H+1)}$$

for large $T$ and fractional Brownian motion,

(6) $$x_i = \sum_{k=1}^{i} z_k$$

where $z_k$ represents fractional Gaussian noise satisfying (see (1.2.1.5-15)),

(7) $$\langle z_j z_k \rangle = \frac{\sigma^2}{2} \left( |j-k+1|^2 H + |j-k-1|^2 H - 2 |j-k|^2 H \right)$$

and the *scaled windowed variance* in (5) is defined by

$$(8) \qquad SD_T^2 = \frac{1}{N/T} \sum_{k=1}^{N/T} \left( \frac{1}{T-1} \sum_{j=1}^{T} \left( x_{T(k-1)+j} - \bar{x}_{T,k} \right)^2 \right)$$

where

$$(9) \qquad \bar{x}_{T,k} = \frac{1}{T} \cdot \left( \sum_{i=T(k-1)+1}^{Tk} x_i \right)$$

In the same paper it is demonstrated, again using (7), that for the fractal Gaussian noise $z$ (7) the *sample variance* defined by

$$(10) \qquad SV_T^2 = \frac{1}{N/T} \sum_{i=1}^{N/T} \left( \frac{\sum_{j=1}^{T} z_{T(i-1)+j}}{T} \right)^2$$

has the average value

$$(11) \qquad \langle SV_T^2 \rangle = \sigma^2 \, T^{2H-2}$$

These theoretical results suggest that the "dispersion methods" are well adapted for extracting the Hurst-exponents if the generation mechanism of the data indeed has fractal properties. However, Raymond and Bassingthwaighte (1999) emphasize that the variance sample method is only adapted for fGn, whereas the scaled windowed variance should be applied to fBm, otherwise the analysis might result in errors. Thus, one must be able to first test to which group the data belongs in order to be able to choose the right method of analysis (Eke et al. 2000).

### 1.2.6 Surrogate data

*"Surrogate data testing attempts to find*
*the least interesting explanation that*
*cannot be ruled out based on the data."*
(Schreiber & Schmitz 1999)

Above we discussed methods for testing for determinsm for instance by checking whether there is predicability in time series or not. This may be quite simple for strongly deterministic signals. However, we may have signals that are heavily cluttered with noise and we whish to know if there are residues of deterministic or nonlinear processes that can be traced in the signal. This problem can be formulated as hypothesis testing. The *null hypothesis $H_0$* might be that the signal is generated by a linear gaussian process (*ARMA, Auto Regression and Moving Average*)

$$(1) \qquad x_n = a_0 + \sum_{i=1}^{M_{AR}} a_i x_{n-i} + \sum_{j=0}^{M_{MA}} b_j \eta_{n-j}$$

where $\eta_j$ designate gaussian independent stochastic variables. The coeffcients $a_i$ and $b_j$ determine the power spectrum and the correlation function of the generated signal which have to agree with that computed from the measurement data. Using several sets of surrogate data one may compute the values of the *estimator* $\lambda$ as sequence of nonlinear parameters $\lambda_i$, the *estimator* $\lambda$ may e.g be the correlation dimension $D_2$, and compare them with that computed from the measurement data, $\lambda_0$. If $\lambda_0$ lies outside, or at the edge of the set $\lambda_i$, we might try to estimate the chance that the null hypothesis is false (see e.g. Kantz & Schreiber 1997). One estimator that has been found to be quite reliable is the prediction error (see section 1.2.3). Thus, in a ("non parametric") statistical test where we use the prediction error as the estimator and we choose the probability, that we might reject $H_0$ though it is in fact true, to be $\alpha$ (typically chosen as 0.05), we have to design a number of $1/\alpha - 1$ sets of surrogate data. If for any of these the prediction error is less than for the original data we *cannot rule out* the null hypothesis given the preset confidence level of $1 - \alpha$. Approximate entropy ApEn discussed in section 1.2.2.1 could also be a candidate for a robus estimator but no systematic investigation seems to have been undertaken regarding this yet.

Demonstraing that the null hypothesis cannot be rejected on the basis of the given evidence does of course not prove that the dynamics is really that of linear gaussian process. The point is rather that the available evidence does not allow us, with the methods at hand, to extract any more subtle structures from the data. For example, few believe that the brain is a linear gaussian process even if the brain waves may have the characteristics of such processes.

A simple method for fabricating surrogate data is the *phase randomization*. For a time series $x_n$ we compute the Fourier coeffcients

$$(2) \qquad c_k = \frac{1}{N} \sum_{n=1}^{N} x_n \cdot e^{-i 2 \pi \frac{k \cdot n}{N}}$$

We then define new coeffcients by multiplying the old coeffcients $c_k$ with complex unit factors having randomly assigned phases $\alpha_k$ in the interval $(0, 2\pi)$:

$$(3) \qquad \bar{c}_k = c_k \cdot e^{i \alpha_k}$$

Applying the inverse transformation to these coeffcients we get a series $x^{(s)}_n$.

$$(4) \qquad x_n^{(s)} = \sum_{k=1}^{N} \bar{c}_k \cdot e^{i 2 \pi \frac{k \cdot n}{N}}$$

This "surrogate data" $x^{(s)}{}_n$ has the same spectrum and correlation function as the original series $x_n$, since from (3)

$$(5) \qquad \left| \bar{c}_k \right| = \left| c_k \right|$$

and the correlation function is obtained as the Fourier inverse of the power spectrum. Thus methods based on spectrum and the correlation function cannot distinguish a series and its phase randomised variant. Phase randomised data has been used to the test the null hypothesis of linear gaussian process. A more general null hypothesis is obtained if we allow for the possiblilty that the signal is a static nonlinear transformation $s(x_n)$ of a linear gaussian process $x_n$. For "nice" forms of the transformation s (i.e. $s$ is invertible and close to identity) we can test this hypothesis by fabricating surrogate data using *Amplitude Adjusted Fourier Transform* (AAFT) (see Schreiber & Schmitz 1999). A very simple procedure is to test for the (somewhat crude) null hypothesis that the data is generated by picking numbers randomly from a set – in this case one creates surrogate data by reshuffling the measurement data in a random way.

Since ARMA processes are time symmetric whereas most nonlinear systems of interest are dissipative one may also use methods for detecting time reversiblilty as a way to to test the null hypothesis of a linear stochastic process. A popular estimator in this case is the 3d order correlation

$$(6) \qquad \phi^{\text{rev}}(k) = \frac{\sum_{n=k+1}^{N} \left| x_n - x_{n-k} \right|^3}{\left( \sum_{n=k+1}^{N} \left| x_n - x_{n-k} \right|^2 \right)^{\frac{3}{2}}}$$

### 1.2.6.1 Diks' estimator

Diks (see Diks 1999) presents a somewhat more advanced method for testing time reversibility which also covers static nonlinear transformations of the linear gaussian processes. The time inversion transformation $P$ is defined by

$$(1) \qquad P(x_1, \dots, x_n) = (x_n, \dots, x_1)$$

The time reversibility is then formulated in terms of the map $P$ and the *reconstruction measure* $\mu$. The measure $\mu$ is defined in the embedding dimension $m$ by the equivalence ($z_n$ are the embedding vectors in dimension $m$)

(2) $$\int_{R^m} f(x)\, d\mu = \frac{1}{N-m+1} \cdot \sum_n f(z_n)$$

for any continuous function $f : R^m \to R$. Diks then formulates the null hypothesis $H_0$ of time reversibility as

(3) $$P\mu = \mu$$

( $P\mu$ is defined by $P\mu(A) = \mu(P^{-1}A)$ for any set $A \in R^m$.) The idea is then to find an estimator which measures the difference between $\mu$ and $P\mu$. For this Diks constructs a metric on the set of measures given by

(4) $$\left(\mu_1, \mu_2\right) \equiv \iint K(r-s)\, d\mu_1(r)\, d\mu_2(s)$$

where the *kernel K* is chosen to be, for some positive number $d$ (called *bandwidth*),

(5) $$K(x) = e^{\frac{-|x|^2}{2 \cdot d}}$$

Now we can define a (squared) difference between $\mu$ and $P\mu$ as

(6) $$Q = (\mu - P\mu, \mu - P\mu) = 2\,(\mu, \mu) - 2\,(\mu, P\mu)$$

which can be written as

(7) $$Q = \iint g(r, s)\, d\mu(r)\, d\mu(s)$$

with

(8) $$g(r, s) = 2\, K(r-s) - 2\, K(r - Ps)$$

This squared difference $Q$ will be the basis for the time reversibillity estimator. Indeed, define the $w$-variables by

(9) $$w_{ij} = g\left(z_i, z_j\right)$$

in terms of the embedding vectors $z_i$. Diks demonstrates that the estimator

$$(10) \qquad S = \frac{\hat{Q}}{\sigma(\hat{Q})}$$

under the null hypothesis plus the assumption that the embedding vectors are statistically independent, has a zero mean and unit variance. Here $\hat{Q}$ is the statistical estimate of $Q$,

$$(11) \qquad \hat{Q} = \frac{1}{\binom{N-T+1}{2}} \cdot \sum_{i=1}^{N-T} \sum_{j=i+T}^{N} w_{ij}$$

($T$ is the usual Theiler correction discussed in section 1.2.2) and $\sigma(\hat{Q})$ is an estimate of its standard deviation,

$$(12) \qquad \sigma(\hat{Q}) = \frac{1}{\binom{N-T+1}{2}} \cdot \sqrt{\sum_{i=1}^{N-T} \sum_{j=i+T}^{N} w_{ij}^{2}}$$

Indeed, under the given assumptions it can be argued that $w_{ij}$ and $w_{kl} (i \neq j, k \neq l)$ are uncorrelated, and that due to time reversibility the values $w_{ij} = \pm a$ have equal probabilities whence $w_{ij}$ (and therefore $\hat{Q}$) has zero mean. Typical parameters used by Diks are $m = 3$, $d = 0.2$ and $T >> m$. One modification (*the block method*) of this procedure in order to reduce the fluctuations of the results is to replace $w_{ij}$ by its average over blocks of the size $l \times l$. An interesting point is that for non reversible series the Diks' estimator $S$ tends to grow with the length of the time series which thus amplifies the discriminatory power of the method. As a rule of thumb, if the null hypothesis is correct then

$$Prob(S > 3) < 0.05.$$

Applying this method to the Wolf sunspot data (using delay = 1, $m = 3$, $d = 0.2$, $T = l = 8$) Diks obtained $S = 5.78$ "which provides strong evidence against the hypothesis that the time series is reversible" (Diks 1999 p. 71). A similar analysis of a financial time series (difference of logarithm of IBM stock price, $\log(x_n) - \log(x_{n-1})$) resulted in $S = 1.55$ which does not rule out reversibility (which is expected from the random walk models).

# 2. Physical models

## 2.1 Human balance

### 2.1.1 Force plate design

Standing and walking involve a complex neurophysiological control system. That neurological or muscular disorder may manifest themselves in abnormal patterns in posture and balance is obvious. Making quantitative measurements of the variations in posture remained a technical challenge however. Systematic studies of human stance seems to have begun only with Karl von Vierordt around 1860. Vierordt affixed a feather on the head of his subjects above of which was a sheet covered with lampblack (fine soot).

When the subject moved, the feather inscribed a trajectory on the sheet. An electronic force plate was constructed by a French professor J Scherrer in 1952. In the 1980's the computer entered the game, used both for data acquisition and analysis. Force was measured with sensors based on pressure gauges or piezo-electric components. The international society of posturography was founded in Amsterdam in 1969 (International Society of Posturography). Posturology seems to be a most active field in countries such as France, Italy and Japan. For some historical tidbits see Gagey & Bizzo (www2001). A good survey of theoretical, experimental and clinical aspects of postural control can be found in Shumway-Cook and Woollacott (2001).

In quiet standing the centre of mass (CM) of the subject makes what seems an erratic bounded trajectory. Suppose the person stands on a rectangular plate supported by force transducers at the corners. The transducers pick up the vertical component of the pressure and we may denote these forces by F1, ..., F4. Knowing the locations of the transducers we can calculate the centre of pressure (COP). Let the $y$-axis be in the anterior-posterior direction, and the $x$-axis along the medial-lateral direction. If sensors are placed at the corners with the co-ordinates ($\pm a, \pm b$) then the co-ordinates of the COP are given by

(1)
$$x = a \cdot \frac{F2 + F3 - F1 - F4}{F1 + F2 + F3 + F4}$$
$$y = b \cdot \frac{F1 + F2 - F3 - F4}{F1 + F2 + F3 + F4}$$

It can be pointed out here that the subtractions of the forces in (1) help also to cancel external perturbations common to all the transcducers. The sum F1 + F2 + F3 + F4 should be almost constant and equal to the weight ($W$) of the person. The accuracy of the transducers are usually given as percentage $p$ of the nominal load ($F_{max}$). Practical measurements seem to indicate that $p\,F_{max}$ is the standard deviation of the error of the transducer (the square root of the variance of the fluctuations around the mean value for a constant load). If we assume that errors from the different transducers are statistically independent we can estimate the errors in the COP co-ordinates to be

(2)
$$\delta x = 2\ a\ \frac{p \cdot F_{max}}{W}$$
$$\delta y = 2\ b\ \frac{p \cdot F_{max}}{W}$$

For instance, precision force transducers may have $p$ = 0.02 % ( = 0.0002). Thus, if we use 100 kg sensors ($F_{max} \approx$ 1000 N) and the person has a body mass of 70 kg, and $a$ = 300 mm, then we find the error (in the sense of standard deviation) $\delta x$ to be of the order of 0.17 mm. Using a 16-bit AD-card interfacing the computer the influence of the AD-errors on the total error becomes in this case negligible ( $p > 2^{-16}$ ). The error of 0.17 mm can be compared to typical amplitudes of the sways which are of the order of $10 - 20$ mm.

### 2.1.2 COP versus CM

In static equilibrium the CM and the COP would lie on the same vertical line (i.e. COP would coincide with the projection of the CM onto the *x-y* plane). This can be illustrated using a simple model, the *inverted pendulum* (Winter at al. 1998), for the anterior-posterior balance. Here we take the vertical direction as the *z*-axis. The pendulum rotates around the ankle joint which we take as the origin in the *y-z* system. Denote by $\boldsymbol{F}$ the total force acting on the foot by the force plate at the point $(-\zeta, \eta)$ which is the COP. Finally we denote the CM co-ordinates by $(y, z)$.

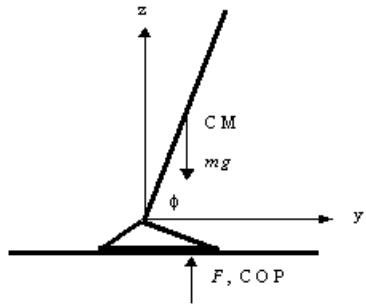

From Newton's equations we obtain (*L* is the the distance between CM and the ankle joint, *I* is the moment of inertia, *m* is the body mass minus feet mass)

(1)
$$m\,\ddot{y} = F_y$$
$$m\,\ddot{z} = F_z - m\,g$$
$$I\,\ddot{\phi} = \eta\,F_z + \zeta\,F_y - m\,g\,L\cos(\phi)$$

The component $F_z$ is the same force as is obtained from the readings of the force transducers. For small deviations around the *z*-axis we may replace $\cos(\phi)$ by $y/L$ and as a first approximation we may also set $F_z = mg$[9], then we obtain from (1)

(2)
$$y - \eta \approx \left( \frac{\zeta}{g} + \frac{I}{m\,g\,L} \right) \cdot \ddot{y}$$

Thus the difference between the projection of the CM and the COP on the *y*-axis is proportional to the second derivative of the CM co-ordinate *y*. We can estimate the difference of $y - \eta$ by approximating the RHS of (2.1-4) as $\dfrac{L}{g} \cdot \ddot{y}$ and using $L \approx 1$ m. If we assume a sway of 5 mm of the CM in 1 second of time we can estimate $\ddot{y}$ to be around 1/100 m/s² and the corresponding difference $y - \eta$ to be around 1 mm. If we take the Fourier transform of (2) we obtain the equation

---

9   Actually the beating of the heart, among other things, causes a fluctuation in the vertical force, typically with an amplitude of the order of 3 - 8 Newtons, which is about the order of 1% of the person's weight *mg*.

(3)
$$\hat{y}(f) = \hat{\eta}(f) \cdot \frac{1}{1 + \dfrac{f^2}{f_c^2}}$$

where $f_c$ is a characteristic frequency of the body given by

(4)
$$f_c = \frac{1}{2\pi} \cdot \sqrt{\frac{g}{\zeta + \dfrac{I}{mL}}}$$

which can be estimated to be typically around 1/3 Hz for adults. From (3) it is apparent that the CM co-ordinate $y$ is a filtered version of the COP co-ordinate $\eta$. This is physically obvious because it takes more energy to move a bigger mass whence vibrations transmitted to a big mass are damped. For slow motions with frequencies $f$ below $f_c$ the co-ordinates $y$ and $\eta$ trace out almost identical curves, but the high frequency components $f > f_c$ are suppressed in the CM $y$ co-ordinate. If we express (3) in the time domain then the smeared nature of the CM co-ordinate becomes even clearer:

(5)
$$y(t) = \pi f_c \cdot \int_{-\infty}^{\infty} e^{-2\pi f_c |t-s|} \eta(s)\, ds$$

Thus, $y(t)$ is $\eta(s)$ "averaged" over a time interval whose characteristic size is typically of the order $1/(2\pi f_c) \sim 0.5$ seconds. Apparently (5) is a *non causal* filter. However, it is equivalent to the solution of the initial value problem given by

(6)
$$y(t) = 2\pi f_c \int_0^t \sinh(2\pi f_s(s-t))\eta(s)\, ds + y(0) \cdot \cosh(2\pi f_s t) +$$
$$\frac{\dot{y}(0)}{2\pi f_s} \cdot \sinh(2\pi f_s t)$$

if $y$ satisfies the condition $(|y(t)| + |\dot{y}(t)|) \cdot \exp(-|t|) \to 0$ as $t \to \pm\infty$. Indeed, given this condition and inserting $\eta$ into the integral (5) using (2) it will yield $y$ after partial integrations.

### 2.1.3 Stiffness hypothesis

An complementary analysis can be formulated in terms of stiffness hypothesis (Winter et al. 1998) which supposes that a deviation by an angle $\alpha$ from the vertical is opposed by muscular torque $K \cdot \alpha$ around the ankle joint, with $K$ as the "spring constant". This must be equal to $\eta F_z + \zeta F_y$ in (2.1.2-1) otherwise there would be a nonzero total torque acting on the foot making it rotate. With $\alpha = \pi/2 - \phi \approx y/L$ we thus obtain

(1)
$$K\frac{y}{L} \approx \eta\, m\, g + \zeta\, m\, \ddot{y} \quad \text{(from stiffness hypothesis)}$$

The last term on the RHS of (1) may be of the order $10^{-4}$ times smaller than first one whence $y$ is quite nearly proportional to $\eta$ and thus in phase with it. In terms of the angle $\alpha$ we further obtain the equation (inverted pendulum with a restoring spring)

(2)
$$\ddot{\alpha} + \left( \frac{K - m\,g\,L}{I} \right) \cdot \alpha = 0$$

which implies a natural frequency of

(3)
$$f_n = \frac{1}{2\,\pi} \sqrt{\frac{K - m\,g\,L}{I}}$$

Accordingly, if the maximum velocity of sway is $v_0$ we can predict the amplitude $y_0$ of the sway to be

(4)
$$y_0 = \frac{v_0}{2\,\pi\,f_n}$$

Conversely, knowing $y_0$ and $v_0$ from measurements we can estimate $f_n$ and the "stiffness constant" $K$. In the cited study they were able to independently measure the COM coordinate $y$ using markers and optical trackers. Then by determining the angle $\alpha \approx y/L$ and the restoring torque $M = mg\eta$ they were able to establish a relation (linear regression typically with $R^2 = 0.954$) between $\alpha$ and $M$ and calculate the stiffness constant from $K = dM/d\alpha$. Average values obtained by (Winter et al., 2001) for the stiffness (anterior/posterior) was about 430 N m/rad per ankle joint. Further evidence for the stiffness- and inverted pendulum modell for the balance mechanism is presented in (Winter at al. 2003). The stiffness modell implies a passive balance control mechanism. Indeed, Winter at al. (2003) found that CM and COP are almost in phase, lagging only by about 4 ms, whereas afferent and efferent latencies are of the order of $50 - 70$ ms and delays in muscle tension are about 70 - 100 ms.

Without the $K$-term the Eq (2) of course describes an unstable situation (the inverted pendulum topples over). Adding the stiffness term $K > mgL$ is the simplest way of stabilising the system (a frictional term would also be needed in order for the system to settle down in the position $\alpha = 0$). Very likely there is a corresponding mechanism in the nature, at least as a linear approximation of a more complicated restoration force. The stiffness model does however not explain the random walk characteristics (Collins & De Luca 1994) of quiet standing. The figure below shows the sways in medial-lateral direction in terms of COP (in mm along the $y$-axis in he graph) in a 15 seconds trial. The dotted line is the average position $mean(y)$.

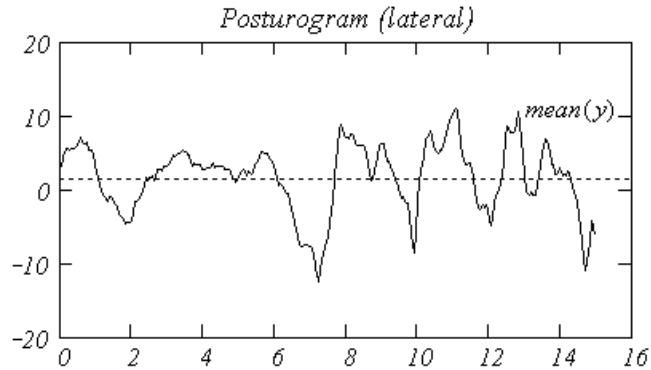

### 2.1.4 Random walk in quiet standing

That there is a random component in the swaying during quiet standing seems to be well established (Collins & De Luca 1994; Duarte & Zatsiorsky 2000), but to what extent this randomness is due to deterministic chaos or some stochastic influence is not settled. Viewing the posturogram as a fractal curve and calculating its Hurst exponent there seems to be a critical value for the size of time interval $\Delta t$ around 1 sec (cmp Eq (1.2.5-12)) such that the Hurst exponent calculated for $\Delta t < 1$ sec tend to be larger than 0.5 while for time intervals $\Delta t > 1$ sec it tend to be smaller than 0.5. (The analysis of posturograms in terms of random walk and fractal properties is called *stabilogram diffusion analysis*, SDF, in the litterature.) This would mean that we have persistence for small time intervals and antipersistence for large time intervals. Indeed, for short time intervals we may suppose that the inertia will tend to keep the body moving in one direction. On the other hand the motion is bounded (otherwise the person topples over) whence for longer time intervals there will be a rebounding of the motion in the opposite direction. Thus, qualitatively the inertia and the boundedness seem to imply the persistence and antipersistence regimes of the swaying. It is simple to extend the Eq (2.1.3-2) with a stochastic term so as to get the random walk signature of the swaying. A somewhat more detailed model (still based on the inverted pendulum) has been presented by Peterka (2000) in terms of PID-control systems known from engineering (see e.g. Khoo (2000) for a physiological context of these methods). Basically it's an ARMA-process. We present the model here in the familiar form of a differential equation (we have dropped a time delay used in the original formulation since it does not seem to affect the results for this model):

$$(1) \qquad I\,\ddot{\alpha} + K_D\,\dot{\alpha} + \left(K_P - mgL\right)\cdot\alpha + K_I\int_0^t \alpha(u)\,du = \frac{1000}{\tau_f}\cdot\int_0^t e^{\frac{-(t-u)}{\tau_f}}\,n(u)\,du$$

The $K$-parameters represent the standard PID-parameters: the damping parameter $K_D$, the stiffness parameter $K_P$ and the integral parameter $K_I$ (which controls the steady-state deviation from the upright position, $\alpha = 0$). On the right hand side we have Gaussian white noise $n(t)$ (uncorrelated) with unit variance and zero mean. The factor of 1000 on the RHS determines the amplitude of the sways, whereas the integration on the RHS acts as a lowpass filter with a characteristic frequency $1/\tau_f$. We quote typical parameter values used by Peterka (2000) and expressed in MKS-units.

| Parameter | Value (MKS) | Explanation |
|:---:|:---:|:---|
| $I$ | 66 | moment of inertia vis-à-vis ankle joints (kg m²) |
| $m$ | 76 | body mass exlusive feet (kg) |
| $K_{\mathrm{D}}$ | 257,8 | damping parameter (N m s rad⁻¹) |
| $K_{\mathrm{P}}$ | 1117 | stiffness parameter (N m rad⁻¹) |
| $K_{\mathrm{I}}$ | 14,3 | integral parameter (N m s⁻¹ rad⁻¹) |
| $L$ | 0,86 | distance from ankle joints to CM (m) |
| $\tau_f$ | 80 | lowpass filter time scale (s) |

The Eq (1) can be conveniently handled numerically if we take the Fourier transform. Thus, one generates the Gaussian noise, takes its Fourier transform, applies the lowpass filtering obtaining $\hat{n}_{\mathrm{filt}}(\omega)$, expresses the Fourier transform of the angle $\hat{\alpha}(\omega)$ in terms of the $\hat{n}_{\mathrm{filt}}(\omega)$, and finally applies the inverse Fourier transform obtaining the angle $\alpha(t)$. Indeed, Eq (1) in the frequency domain is:

(2) $$\hat{\alpha}(\omega) = \frac{1}{K(\omega)} \cdot \frac{1000}{1 + i\,\omega\,\tau_f} \cdot \hat{n}(\omega)$$

where

(3) $$K(\omega) = -I \cdot \omega^2 + K_{\mathrm{D}}\,i\,\omega + K_{\mathrm{P}} - m\,g\,L + K_{\mathrm{I}} \cdot \frac{1}{i\,\omega}$$

The CM-sway in mm will be given by

$$y = 1000 \cdot L \cdot \alpha$$

Using Eq (2.1.2-3) we obtain the Fourier transform of the COP co-ordinate $\eta$ from that of $\alpha$. The figure below shows a simulated example of COP trajectory $\eta(t)$ (solid line) and CM trajectory $y(t)$ (dotted line) in millimeters along the vertical axis and with time (seconds) along the horizontal axis. This result was obtained by the Fourier transform method described above using the time step $\Delta t = 0.01$ s. (An artefact of the Fourier method is that the trajectory always ends and starts at the same value due to the periodicity of the sinus and cosinus functions in the Fourier expansion. For unbiased statistics one may use only parts of the simulated curve.)

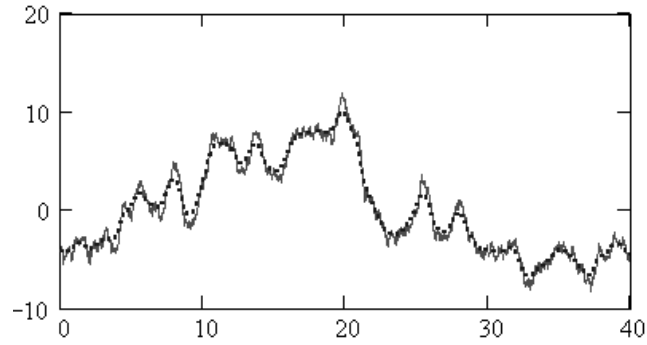

From (2) we may directly obtain the spectrum $\left|\hat{y}(\omega)\right|^2$ for the CM co-ordinate $y$. With the above choice of parameters there is evidence for a "critical point" around the natural frequency

$$(4) \qquad f_n = \frac{1}{2\pi}\sqrt{\frac{K_P - mgL}{I}} \approx 0.42 \,\text{Hz}$$

where the power curve changes its slope (in a log-log graph). In the figure (log-log plot) below the solid line describes the theoretical spectrum for the CM trajectory while the dotted line describes the spectrum of the COP trajectory obtained from

$$(5) \qquad \left|\hat{\eta}(\omega)\right|^2 = \left(1 + \frac{\omega^2}{\omega_c^2}\right)^2 \cdot \left|\hat{y}(\omega)\right|^2$$

where

$$(6) \qquad \omega_c = 2\pi f_c = \sqrt{\frac{mgL}{I}}$$

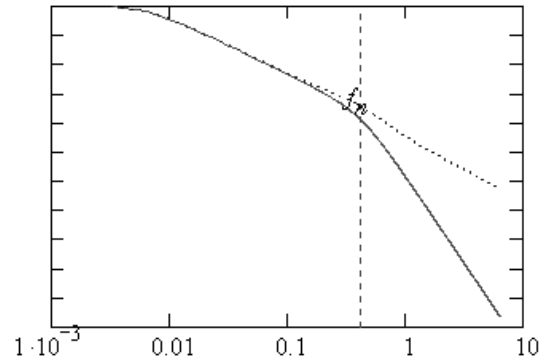

There is much less dramatic change in the spectrum of the COP co-ordinate. Still if we compute the slope $-\beta$ of its spectrum (in the log-log graph) for the section $f < f_n$ ($\beta_l$, for long time) and the section $f > f_n$ ($\beta_s$, for short time) we obtain in this example

$$\beta_l \approx 1.685$$

$$\beta_s \approx 2.249$$

with the corresponding Hurst exponents (see Eq (1.2.5-10))

$$H_l \approx 0.343$$

$$H_s \approx 0.624$$

Thus, the model seems to be able to describe the observed persistence for short time intervals ($H_s > 0.5$) and the antipersistence for long time intervals ($H_l < 0.5$). Whether such a PID-control scheme is realised in nature is an open issue. The nervous system may rely on inputs from e.g. muscle spindles, joint receptors, vestibular sensors in the inner ear (semicircular canals, otolithic organs), gravity receptors in the trunk, plus visual cues. Also it has to be pointed out that the inverted pendulum model is only a first approximation. Especially elder people tend to control the stance using the so-called "hip strategy"; i.e., shifting the position of the CM by bending the body.

### 2.1.5 Disgression: Random walk with relaxation

Apparent different "scaling regimes" may seem to result when one combines random walk with a friction term as in the following simple model (see e.g. Hastings & Sugihara 1993) which is a Langevin-equation on differential form:

(1) $$dx(t) = -r\, x(t)\, dt + dB(t)$$

Here the increment $dB$ represents uncorrelated noise with zero mean. The formal solution with[10] $x(0) = 0$ can be written as

---

10  This condition means that the averages below are taken over trajectories that pass through $x = 0$ at $t = 0$.

(2)         $$x(t) = \int_0^t e^{r(s-t)} dB(s)$$

We have especially

(3)
$$x(t + \Delta t) = e^{-r\Delta t} x(t) + \Delta W(t)$$
$$\Delta W(t) = e^{-r\Delta t} \int_t^{t+\Delta t} e^{r(s-t)} dB(s)$$

with $\langle \Delta W(t) \rangle = 0$ and $\langle \Delta W(t) x(t) \rangle = 0$ by assumption. For a long time $t$ enough the process becomes stationary since from

(4)         $$\langle x(t + \Delta t)^2 \rangle = e^{-2r\Delta t} \cdot \langle x(t)^2 \rangle + \langle \Delta W \rangle^2$$

we obtain

(5)         $$\langle x(t + n\Delta t)^2 \rangle = e^{-2rn\Delta t} \cdot \langle x(t)^2 \rangle + \langle \Delta W \rangle^2 \cdot \frac{1 - e^{-2rn\Delta t}}{1 - e^{-2r\Delta t}}$$

which approaches (introducing a parameter $\sigma$)

(6)         $$\frac{\langle \Delta W \rangle^2}{1 - e^{-2r\Delta t}} \equiv \sigma^2$$

as $n$ goes to infinity.

Thus for large times $t$ we may set

(7)         $$\langle x(t)^2 \rangle = \sigma^2$$

Using Eqs (7) and (3) we can finally show that (for large times $t$)

(8)         $$\langle (x(t+\Delta t) - x(t))^2 \rangle = 2\sigma^2 \cdot (1 - e^{-r\Delta t})$$

Compare this with the definition of the Hurst exponent $H$ in Eq (1.2.5-12). Apparently, for $\Delta t \ll 1/r$ we have $H \sim 0.5$; for $\Delta t \sim 1/r$ one might obtain a value $0 < H < 0.5$, and finally for $\Delta t \gg 1/r$ the Hurst exponent $H$ would be concluded to be close to zero. Thus we may expect different "scaling regimes" if one calculates the Hurst exponent from the data without regard to whether there is a real scaling or not. In this model there is a true scaling regime only for short time intervals. We can also look at the situation from the frequency point of view. Eq (3) implies that the correlation $C(\tau)$ is proportional to exp(- $r$ |$\tau$|) whose Fourier transform is

$$(9) \qquad \int_{-\infty}^{\infty} e^{-i2\pi f \tau} e^{-r|\tau|} d\tau = \frac{2\,r}{r^2 + 4\,\pi^2\,f^2}$$

Thus, for small frequencies $f$ (long times) the spectrum is almost constant ($\beta = 0$ in which case the Hurst exponent is undefined), whereas for large frequencies (short times) the power spectrum approaches $1/f^2$. If we calculate the correlation $\rho$ defined by Eq (1.2.5-11) we obtain

$$\rho = -\frac{1}{2} \cdot \left(1 - e^{-r\Delta t}\right)$$

which implies anti-persistence for all time scales $\Delta t$.

### 2.1.6 Some Parameters Used in Posturography

Using the data of the COP-trajectory one can calculate a number of parameters with the intent characterizing "good" or "poor" balance. Broadly speaking we have three groups of parameters (Chiari et al. 2002). The (*space-*)*time-domain* parameters, the *frequency-domain* parameters, and the *stochastic-nonlinear* parameters. Classical time-domain parameters are the length of the COP-trajectory and the area covered by the COP-trajectory. The "length"

$$L = \sum_k \sqrt{\left(x[k] - x[k-1]\right)^2 + \left(y[k] - y[k-1]\right)^2}$$

naturally depends - considering the raggy and fractal characteristics of the trajectory - on the sampling time interval $\Delta t = t_k - t_{k-1}$. For small intervals noise will give a spurious contribution to the length. A convenient choice is using $\Delta t = 0.2$ sec for calculating the length which accords with the old standard of using a sampling rate of 5 Hz. With this value the noise is effectively suppressed.

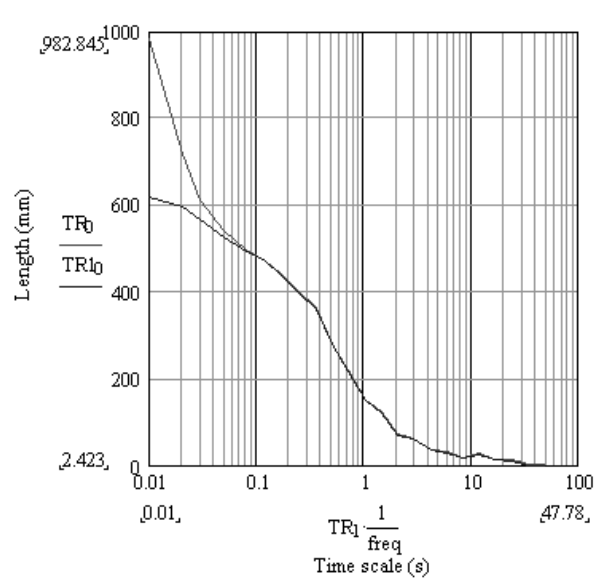

*Length as a function of the time interval. Red curve based on raw data sampled at 1000 Hz and replaced by 10-point average data with 100 samples per second, blue curve based on the same data but low-pass filtered with cut-off frequency at 20 Hz. Data from a 52 second test eyes open.*

A standard measure of area is obtained by calculating the area of the 90%-confidence ellipse for the $x$ and $y$-data of the COP-trajectory. Of interest is also the inclination of the ellipse. Other "obvious" parameters are the mean velocities and the standard deviations of the velocities in the A/P and M/L-directions. Among frequency-domain parameters we have mean the frequencies for the spectrum of the A/P and M/L motion. Examples of stochastic parameters are the Hurst parameters and the power spectrum exponent $\beta$ discussed in section 2.1.4. A higher β-value is generally associated with a smoother curve since the high-frequency part is suppressed. In a sense a smaller β-value signifies a more "complex" curve. These types of parameters are of increasing significance since they seem to be able to capture features that are hard to detect using the time-domain parameters (Thurner, Mittermaier and Ehrenberger 2002). The quoted study reports e.g. that data for groups of elderly show on the average significantly higher β-values than for a group of young people. We have also implemented entropy parameters such as approximate entropy but it is too early to tell whether they are useful in this context.

## 2.2 Analysis of EMG (ElectroMyoGram)

### 2.2.1 Neuromuscular Basics

That there is connection between electricity and muscle activity was discovered by Luiggi Galvani already in 1781. Jan Swammerdam hade made similar experiments more than a century earlier (1664) but it is unclear whether he was aware of the role of the electricity in his experiments. A more detailed study of the phenomenon had to wait for developments in anatomy, microscopy and electronics (for historical tidbits see e.g. the introductory chapter in Malmivuo and Plonsey 1995). Skeletal muscles are composed of muscle fibres controlled by the somatomotor nervous system. The muscles exert force when the fibres are contracting. The contraction is effected by the action potentials (AP), descended from an alpha-motoneuron, which generate muscle fibre action potentials,

MAP, which travels in both directions from the motor end platea along the fibres. The action potential along the motoneuron dendrites travels much faster, around 60 m/s, than the muscle action potential along the fibre which may travel with a velocity around 4 m/s. An alpha-motoneuron may control anything from only a few muscle fibres to thousands of muscle fibres (when only gross control of the movements is needed). The alpha-motoneuron, its junctions and the muscle fibres it innervates, are collectively called a *motor unit*, MU. Motoneurons that end up on the same muscle constitute a *motor pool*. If electrodes are inserted into the muscle they may record a sum of the fibre action potentials, whose waveform is called the *motor unit action potential*, MUAP. For a sustained contraction the alpha-motoneuron has to emit a train of action potentials, causing a similar train of motor unit action potentials, MUAPT. The intensity of the contractions is determined by the *firing rate* of the alpha-motoneuron. A few random elements are involved in the neuromotoric process. For example, the release of the acetylcholine (ACh) packets at the neuromuscular junctions is a random discharge causing a random excitation of the muscle fibres called *jitter*. Generally the fibers in a motor unit are activated simultaneously (save for the shifting delays due to varying lengths of the dendritic connections from the motoneuron), but sometimes some of the fibers fail to fire which also adds to the randomness of the total potential.

For surface muscles it is possible to detect a gross electrical activity (the recorded signal is called a *surface electromyogram*, sEMG) of the muscles using surface electrodes. This technique was pioneered by H Piper back in 1912. The amplitude of the muscular action potential is of the order from a few μV to 1000 μV. The frequency content of the MAP is in the range from a few Hz to 10 kHz, but due to filtering by skin, fat, etc., the sEMG is believed to contain relevant information only up to about 500 Hz. The part below 20 Hz is usually filtered out because it may contain motion artefacts and traces from the electro-cardiac potentials. Surface EMG usually records the superposition of MAPs from many motor units and may also contain crosstalk from other muscles. Using arrays of electrodes it may be possible to isolate individual motor unit components. For a standard reference on electromyography consult Basmaijan and de Luca (1985). Recent reviews, together with many references, on the advances in EMG modelling and analysis can be found in the special issues of *Medical Engineering and Physics* (Vol. 22, issues 6 - 7, 1999) and *Journal of Electromyography and Kinesiology* (Vol. 10, 2000).

The most common parameters used in the study of EMG is the *integrated rectified EMG* (iEMG), the *Random Mean Square value* (RMS), the *average frequency* and the *median frequency*. The two first parameters are defined for a given period $T$ and EMG-signal $m$ as:

$$iEMG(t) = \int_{t}^{t+T} |m(s)| \, ds$$

(1)

$$RMS(t) = \sqrt{\frac{1}{T} \cdot \int_{t}^{t+T} |m(s)|^2 \, ds}$$

The average frequency $f_{av}$ is defined with respect to the spectrum of the EMG-signal as

(2) $$f_{av} = \frac{\int f \left| \hat{m}(f) \right|^2 df}{\int \left| \hat{m}(f) \right|^2 df}$$

The median frequency $f_m$ is defined such that

(3) $$\int_0^{f_m} f \left| \hat{m}(f) \right|^2 df = \frac{1}{2} \cdot \int f \left| \hat{m}(f) \right|^2 df$$

One can also trace the time evolution of the frequency parameters by dividing the EMG times series into subintervals and calculate the spectrum for each subinterval and then use Eqs (2 - 3).

### 2.2.2 Modelling EMG

A MAP travelling along a fibre with a velocity $v$ (taken to be aligned with the $x$-axis) may be described by a "wave-function"

(1) $$g(x - vt)$$

which describes the voltage amplitude of a travelling wave at the time $t$ and the point $x$. Thus, the signal $h(t)$, recorded at a fixed point $x_0$, as a function of time $t$ is given by

(2) $$h(t) = g(x_0 - vt)$$

If we have a train of MAPs arriving at times $t_1, t_2, ..., t_n$ we can write the signal at $x_0$ as a sum of pulses (MUAPT)

(3) $$u(t) = \sum_{i=1}^{n} h(t - t_i)$$

It is generally believed that the interpulse intervals (IPI) $\tau_i = t_i - t_{i-1}$ are generated by a stochastic process. The average number of pulses $\lambda$ per time is termed the *firing rate*. If we take the Fourier transform of (3) we obtain

(4) $$\hat{u}(f) = \sum_{k=1}^{n} e^{-i2\pi f t_k} \cdot \hat{h}(f)$$

If we further take the absolute square of (4) we get for the power spectrum

(5) $$|\hat{u}(f)|^2 = |\hat{T}(f)|^2 \cdot |\hat{h}(f)|^2$$

with the transfer function power spectrum given by

(6) $$|T(f)|^2 = n + 2 \sum_{j>k} \cos\left(2\,\pi\,f\left(t_j - t_k\right)\right)$$

If the interpulse intervals $\tau_k$ are independent random variables with probability density function (pdf) $p(\tau)$ then the *statistical average* of the transfer function power spectrum can be expressed as

(7) $$n + 2 \sum_{k=1}^{n-1} (n-k)\cos(k\,\alpha)\left|\hat{p}(f)\right|^k$$

where $\hat{p}(f) = |\hat{p}(f)| e^{i\alpha}$ is the Fourier transform of the probability density. In the limit of $n \to \infty$ Eq (7) can be simplified by carrying out the summation (reduced to a simple geometric sum) and taking the average over the number of pulses $n$:

(7*) $$\frac{|\hat{T}(f)|^2}{n} = \frac{1 - |\hat{p}(f)|^2}{|1 - \hat{p}(f)|^2} = \frac{1 - |\hat{p}(f)|^2}{1 - 2\,|\hat{p}(f)|\cdot\cos(\alpha) + |\hat{p}(f)|^2}$$

(this essentially the same expression as that given by Basmaijan and de Luca (1985) p. 75). We consider an example where the interpulse intervals are approximated by a Gaussian pdf

(8) $$p(\tau) = \frac{1}{\sqrt{2\,\pi}\,\sigma_\tau} e^{\frac{-(\tau - \tau_m)^2}{2\sigma_\tau^2}}$$

Its Fourier transform becomes

(9) $$\hat{p}(f) = e^{-i2\,\pi\,f\,\tau_m} \cdot e^{-2\,\pi^2\,f^2\,\sigma_\tau^2}$$

Computing the average over the $n$ pulses we obtain in this case from (7) the factor

(10) $$1 + 2 \sum_{k=1}^{n-1} \left(1 - \frac{k}{n}\right)\cos\left(2\,\pi\,f\,k\,\tau_m\right) e^{-k\,2\,\pi^2\,f^2\,\sigma_\tau^2}$$

which apparently acts as a high-pass filter modulated with a frequency $1/\tau_m$. An example of some physiologically realistic values could be $\tau_m = 60$ ms and $\sigma_\tau = 12$ ms. Eq (10) is in qualitative agreement with interpulse interval data (see Basmaijan and de Luca 1985 p. 78) and reaches a practically flat spectrum for $f > 2/\tau_m$ ($\approx 33$ Hz in the present example).

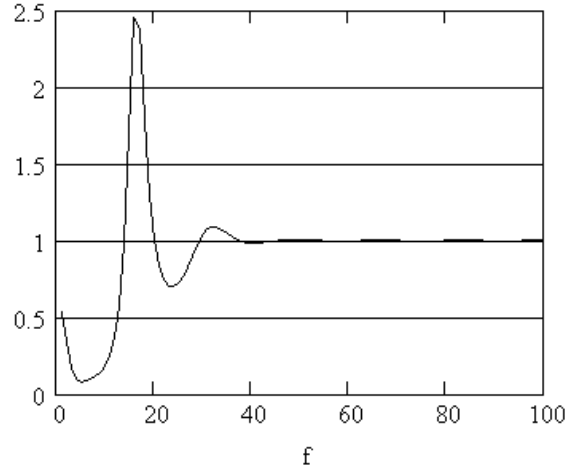

The size of the first peak is sensitive to the size of the standard deviation . Indeed, suppose the interpulse intervals are constant ( ), $_i = _m$ (i = 1, 2, ...), then (10) will approach a sum of peaks at the points k/ $_m$. This follows from Poisson's summation formula (see e.g. Saichev and Woyczyñski 1997 p. 271) which in this context reads

(11) $$\sum_k e^{-ik2\pi f\tau} = \sum_k \delta(f\tau - k)$$

where $\delta(x)$ is the Dirac delta function. On the other extreme side (uniform distribution), if $\sigma_\tau \to \infty$ then (10) approaches the constant function 1.

In sEMG the electrodes will detect MUAPT from several muscle fibers and we have therefore to consider a sum of MUAPTs (3)

(12) $$m(t) = \sum_{p=1}^{N} u^{(p)}(t)$$

Also the parameters characterising the interpulse intervals may change with time due e.g. to fatigue. A general observation is that the conduction velocity $v$ of the fibres decreases with fatigue. The decrease in the conduction velocity is thought to be linked to the decreasing pH-value in the fatiguing muscle. From Eq (3) we obtain that

(13) $$\left|\hat{h}(f)\right| = \frac{1}{v} \cdot \left|\hat{g}\left(\frac{f}{v}\right)\right|$$

which suggests that the power spectrum is shifted toward the lower end when the velocity $v$ decreases. It is indeed observed that during fatigue (at constant force level) the *median* and the *average frequencies* of the power spectrum tend to *decrease*. Indeed, if we compute the *average* frequency of the spectrum (ignoring $T(f)$ for the moment) we obtain

(14)
$$\frac{\int f \left|\hat{h}(f)\right|^2 df}{\int \left|\hat{h}(f)\right|^2 df} \propto v$$

according to (13), which predicts a decrease of the average frequency due to decreasing velocity $v$. This is because the "energy" (the denominator in (14)), the integrated power, is proportional to $1/v$ and thus increases with decreasing velocity $v$. But it is also found that the *firing rate $\lambda$ decreases* with fatigue, perhaps due to adaptation. The firing rate affects the factor (7) since the number of pulse $n$ is given by $n = \lambda t$ where $t$ is the time of measurement. Thus, averaging over the time $t$ we get the factor $\lambda$ and we would therefore predict the *energy* to change as $\lambda / v$.

The combined observed effect is that the energy tends to *increase* with fatigue for sEMG, but to *decrease* for EMG measured with indwelling electrodes. The explanation may be that in the case of the sEMG the shift of the spectrum toward the lower end means that a larger part of the energy will reside inside its bandpass interval, determined by the skin and tissue and the distance between the electrodes, enough to offset the decrease due to the lowered firing rate (see Basmaijan and de Luca 1985 p. 92 – 94, and ch. 8 on muscle fatigue). For indwelling electrodes the bandwidth is much larger and there is no similar effect of a "shadowed energy coming into light".

It is common to measure sEMG using *bipolar electrodes*; that is, with two detecting surfaces. Suppose the electrodes are aligned along the axis of the muscle fibres with a separation $d$. The potential difference between the electrodes due to a travelling wave (1) would be

(15)
$$h(t) = g(x + d - vt) - g(x - vt)$$

whose Fourier transform is

(16)
$$\hat{h}(f) = e^{-i 2 \pi f \frac{x}{v}} \cdot \left(1 - e^{-i 2 \pi f \frac{d}{v}}\right) \cdot \hat{g}\left(\frac{-f}{v}\right)$$

Taking the squared magnitude of (16) we obtain

(17)
$$\left|\hat{h}(f)\right|^2 = \left\{2 \, \sin\left(\frac{\pi f d}{v}\right)\right\}^2 \cdot \left|\frac{1}{v} \cdot \hat{g}\left(\frac{f}{v}\right)\right|^2$$

Thus, the spectrum is modulated by a geometrical sinus-factor with *pass frequencies* given by

(18)
$$f = \left( n + \frac{1}{2} \right) \cdot \frac{v}{d}$$

and the *cancellation frequencies* given by

(19)
$$f = n \cdot \frac{v}{d}$$

for $n = 1, 2, 3, \ldots$ For bipolar surface electrodes we may have $d = 0.01$ m, which for $v = 4$ m/s gives a cancellation frequency ($n = 1$) of 400 Hz. For needle electrodes we may have a separation around $d = 5 \cdot 10^{-4}$ m which gives a corresponding cancellation frequency of 8000 Hz. In careful measurements the dip in the spectrum around the cancellation frequency $v/d$ can be observed and used for determining the conduction velocity $v$.

An interesting observation is that there seems to be a kind of a natural background oscillations of the muscles at constant-force contractions.

"The fluctuations in force which are ever present during attempted constant-force contractions are a manifestation of low frequency oscillations which are inherent in the firing rates of the motor units. The dominant frequency of this oscillation is approximately 1.5 Hz. The source of this oscillation has not been identified yet" (Basmaijan and de Luca 1985 p. 152).

Basmajian & de Luca notes later that this phenomenon may perhaps be related to another property of the muscle. If the muscle is exercised with a sinusoidal varying force at isometric conditions one may estimate the transfer function of the mechano-electric system between the integrated rectified EMG and the force. One quoted approximate result for the transfer function is (from Basmaijan and de Luca 1985, p. 197) which refers to results by Soechting and Roberts 1975)

(20)
$$\frac{1}{\left( i \, 2 \, \pi \, f + 5 \, \pi \right)^2}$$

which thus acts as low pass filter with the cutoff frequency of 2.5 Hz which is of the same order as the frequency of the "natural oscillation" of the muscle.

### 2.2.2.1 Model of IAP

A popular mathematical model for the *intracellular action potential* (IAP) is given as a function of space co-ordinate by (see e.g. Stegeman et al. 2000)

(1)
$$IAP(x) = A \, x^3 \, e^{-\lambda x} \cdot \Theta(x) - B$$

with parameters $A$, $B$ and $\lambda$ which can be chosen so that (1) fits the observed potentials. (The Heviside function $\Theta(x)$ on the RHS makes the first term zero for x < 0.) From the action potential (1) we may also derive the corresponding transmembrane current $I_m$. Indeed, the classical *cable equation*

$$(2) \qquad I_m = C_m \frac{\partial \phi}{\partial t} + I_{Ion} = \frac{\sigma a}{2} \cdot \frac{\partial^2 \phi}{\partial x^2}$$

(see e.g. Keener & Sneyd 1998, ch. 8) for an intensive treatment, and for a more extensive treatement, see (Koch 1999), a book which is largely devoted to the applications of the cable equation to physiology) goes back to works by Lord Kelvin (1859) and Oliver Heaviside (1876) on the theory of the transatlantic cable and was imported into physiology foremost by Hodgkin and Rushton (1946). Though Eq (2) is a relatively simple consequence of the basic "laws" of Ohm and Kirchoff in the theory of electricity it has profound consequences.

In the present context $\phi$ in Eq (2) is the potential difference across the membrane, $C_m$ is the capacitance per unit area of the membrane, $I_m$ is the total transmembrane current densisty, $I_{Ion}$ is the ohmic part of the transmembrane current density, $\sigma$ is the electrical conductivity along the fibre axis, and $a$ is the radius of the fibre. Identitfying $I_m$ with IAP above we obtain that $I_m$ is proportional to the second derivative of the IAP potential,

$$(3) \qquad I_m(x) = \frac{\sigma a}{2} \cdot \frac{d^2 IAP(x)}{dx^2}$$

where $a$ is the radius of the fibre and $\sigma$ is the electrical conductivity along the fibre axis (which here coincides with the x-axis). Since the extracellular space generally have quite high conductivity the potential is nearly *constant on the outer side* of the membrane. It is therefore the current density (3) which carries the information about the potential changes *in the cell* to the outside world. This is described in the next section on the volume conductor model.

The cable equation (2) allows for pulse-like waves (action potentials) in the fibre as was shown in the pioneering works by Hodgkin, Huxely and Katz (1954). By taking into account the active ion pumps of the cell, the $I_{Ion}$ will be described as a nonlinear function of the potential. This in turn is the mathematical basis for the existence of the pulse-like solutions. It may be fitting here to quote the words by Keener and Sneyd (p. 121) on the Hodgkin-Huxley-Katz contribution: "In biology, where quantitatively predicitve theories are rare, this works stands out as one of the most successful combinations of experiment and theory". However, some of the basic tenets of the ion pump theory have been challenged by G Ling (2001) and G H Pollack (2001) on both experimental and theoretical grounds.

### 2.2.2.2 The volume conductor model

Consider first the simple case of a homogenic resistive medium with the current density $\boldsymbol{J}$ and the potential $\phi$ related by the ohmic relation

(1) $\qquad -\sigma \nabla \phi = \boldsymbol{J}$

where $\sigma$ is the conductivity (Eq (1) is essentially a continuum rendering of "Ohm's law" $U = RI$ ). Taking the divergence of Eq (1) and assuming $\sigma$ to be constant we obtain

(2) $\qquad -\sigma \nabla^2 \phi = \nabla \cdot \boldsymbol{J} = I_S$

The last equality in Eq (2) defines the *current source $I_S$*. In order to elucidate the meaning of the current source term we integrate it over a volume $V$ bounded by a surface $S$ using Gauss' divergence theorem

(3) $\qquad \int_V \nabla \cdot \boldsymbol{J} \, dV = \int_S \boldsymbol{J} \cdot d\boldsymbol{S}$

Suppose $S$ is the surface bounding a muscle fibre, then the RHS of Eq (3) gives the current flowing through the surface (the transmembrane current). Comparing Eqs (2) of this and the previous sections we infer that for a cylindrical fiber of radius $a$ we have

$\qquad I_S = \dfrac{2}{a} I_m$

Muscle fibers are however not electrically isotropic. The conductivity along the axis of the fibre is about five times the radial conductivity. Typical values are given in the table below (quoted from Roeleveld et al. 1997)

| Conductivity $\sigma$ (1/Ohm·m) | Tissue |
|---|---|
| 0,10 | Muscle, radial ($\sigma_r$) |
| 0,50 | Muscle, axial ($\sigma_z$) |
| 0,05 | Fat (isotropic) |
| 1,00 | Skin (isotropic) |

The Eq (2) is simple to adapt to the case of the anisotropic muscle. In terms of cylindrical co-ordinates the Poisson equation becomes (the fiber axis along the *z*-axis)

(4) $\qquad \sigma_r \left( \dfrac{1}{r^2} \dfrac{\partial^2 \phi}{\partial \varphi^2} + \dfrac{1}{r} \dfrac{\partial}{\partial r} \left( r \dfrac{\partial \phi}{\partial r} \right) \right) + \sigma_z \dfrac{\partial^2 \phi}{\partial z^2} = -I_m$

In one of the oldest models, described e.g. in Stegeman et al. (2000), one considers the case of a fibre in the muscle. The muscle tissue is assumed to be a volume conductor of infinite extent bounded by an infinite plane $\Gamma$, representing the skin, which is taken to be parallel with the fibre axis. The *boundary condition* for (4) is then that no current passes through the skin,

(5)     $\sigma_r\,\boldsymbol{n}\cdot\nabla\,\phi=0$
at the plane $\Gamma$

where $\boldsymbol{n}$ denotes the normal of the plane. Suppose that the plane $\Gamma$ is at a distance $d/2$ from the fibre. Now it is easy to construct a solution for (4 - 5) from the solutions $\phi_0$ in the case of an unrestricted infinite volume conductor using the mirroring method. Indeed, if we add two such solutions, one with the fibre on the right side of the plane, the other with the fiber on the mirror side of the plane,

(6)     $\phi(\boldsymbol{r})=\phi_0(\boldsymbol{r})+\phi_0(\boldsymbol{r}-\boldsymbol{d})$

then it can be shown that it satisfies the boundary condition (5). Especially on the skin (the plane $\Gamma$) we obtain

(7)     $\phi(\boldsymbol{r})=2\,\phi_0(\boldsymbol{r})$
for $\boldsymbol{r}\in\Gamma$

These results are based on the fact that in the isotropic case $\phi_0$ is a simple sum or integral of $1/|\boldsymbol{r}|$-potentials (which all satisfy Eq (2) in the region where $I_s=0$). The solution in anisotropic case can be obtained from the isotropic solution by rescaling the $z$-variable as $\sqrt{\dfrac{\sigma_r}{\sigma_z}}\cdot z$. This leads finally to the expression

(8)     $\phi(r,z)=\dfrac{2}{4\,\pi\,\sigma_r}\displaystyle\int_0^L\dfrac{2\,\pi\,a\,I_m(s)}{\sqrt{r^2\,\dfrac{\sigma_r}{\sigma_z}+(z-s)^2}}\,ds$

for the potential at the skin due to a MAP in a fibre of length $L$ and radius $a$. The next step is to add more realistic details to the model by considering the fatty tissue and the properties of the skin.

### 2.2.2.3 A four-layer model

We will describe a simple four-layer model (Farina & Rainoldi 1999) which gives an idea how the tissue affects the current generated signal in the muscle. In this model we have a sandwich consisting of a muscle layer (a), a fat layer (b), and finally a skin layer (c) which interfaces with the air (d). The y-axis is taken to be orthogonal to the sections, the x-axis is along the plane orthogonal to the direction of the muscle, whereas the z-axis is taken to be along the muscle.

All the layers are supposed to be isotropic except for the muscle layer which has different conductivities in the transversal (x-axis) and the axial directions (z-axis) as pointed out int the previous section.

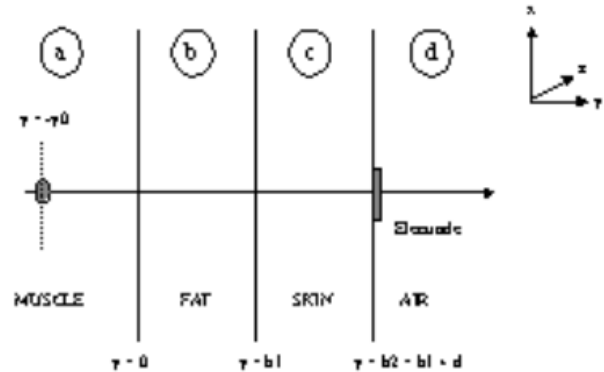

The current source is supposed to be a unit point source located at $(0, -y_0, 0)$

(1) $\qquad I_m(x, y, z) = \delta(x)\delta(y + y_0)\delta(z)$

The solution for a general current source can then be obtained by summing solutions for point sources. This is an example of the methods of the *Green's function* in mathematical physics (Wallace 1984). If we have some linear differential equation

(2) $\qquad \Lambda\phi(\boldsymbol{r}) = \rho(\boldsymbol{r})$

then a Green's function $G(\boldsymbol{r})$ is a solution of the point source equation

(3) $\qquad \Lambda G(\boldsymbol{r}) = \delta(\boldsymbol{r})$

and the solution to (2) can then be obtained from

(4) $\qquad \phi(\boldsymbol{r}) = \int G(\boldsymbol{r} - \boldsymbol{s})\rho(\boldsymbol{s})\, d\boldsymbol{s}$

In signal processing the *transfer function* – the response $H$ to a delta-signal input – is the counterpart of the Green's function. Following Farina and Rainoldi (1999) we will study the solution of (2) in term of the Fourier transformed field

(5) $\qquad \tilde{\phi}(k_x, y, k_z) = \int\int e^{-ik_x x - ik_z z}\phi(x, y, z)\, dx\, dz$

The Eq (4) of the previous section then becomes in the present problem, with the unit current source (1), the following set of equations ($\sigma_{ax} = \sigma_{ay} \neq \sigma_{az}$):

(6)
$$\left(\partial_y^2 - k_{ay}^2\right)\tilde{\phi}(y) = \frac{1}{\sigma_{ay}}\delta(y + y_0)$$
$$k_{ay} = \sqrt{k_x^2 + \frac{\sigma_{az}}{\sigma_{ay}}k_z^2}$$

for layer (a), and for the other layers ($\sigma_x = \sigma_y = \sigma_z$) we get similar equations

(7)
$$\left(\partial_y^2 - k_y^2\right)\tilde{\phi}(y) = 0$$
$$k_y = \sqrt{k_x^2 + k_z^2}$$

whose solutions are composed of simple exponential functions $e^{\pm ky}$ while in the layer (a) we have an "inhomogenic" solution of the type $e^{-k|y + y0|}$ plus the homogenic ones $e^{\pm ky}$. The boundary conditions are that the field, and the current density $\boldsymbol{J}$ (Eq (1) in the previous section), should be continuous at the layer interfaces,

(8)
$$\left(\sigma \nabla \phi\right)_- = \left(\sigma \nabla \phi\right)_+$$
$$\left(\phi\right)_- = \left(\phi\right)_+$$

where "-" ("+") means the left (right) side limit value at a layer interface. Together with the requirement that $\phi(y)$ goes to zero as $y \to \pm\infty$, these equations and conditions determine an unique solution. In fact, we have already used (8) in Eqs (6) and (7) for the $x$ and $z$ directions when using the fact that $k_x$ and $k_z$ do not change going from one layer to another for solutions having an $x$- and a $z$-dependence of the form $e^{k_x x + k_z z}$. The general solution for each layer can be written

(9)
$$\tilde{\phi}_a = \frac{1}{2\,k_{ay}}e^{-k_{ay}|y + y_0|} + u_a e^{k_{ay}y} \quad (y < 0)$$

$$\tilde{\phi}_b = u_b e^{k_y y} + v_b e^{-k_y y}$$

$$\tilde{\phi}_c = u_c e^{k_y y} + v_c e^{-k_y y}$$

$$\tilde{\phi}_d = v_d e^{-k_y y} \quad (y > h_1 + d)$$

with the $u$- and $v$-parameters to be determined by the boundary conditions (8). These conditions result in a matrix equation

$$(10) \qquad K \begin{pmatrix} \dfrac{1}{k_{ay}} e^{-k_{ay} y_0} + u_a \\[2mm] -\dfrac{\sigma_{ay}}{2} e^{-k_{ay} y_0} + u_a k_{ay} \sigma_{ay} \end{pmatrix} = \begin{pmatrix} v_d e^{-k_y(h_1+d)} \\ 0 \end{pmatrix}$$

where the matrix $K$ is given by

$$(11) \qquad K = \dfrac{1}{\sigma_{by} \sigma_{cy} k^2} \begin{pmatrix} k^2 D & k B \\ k^3 C & k^2 A \end{pmatrix}$$

The functions $A$, $B$, $C$ and $D$ are in turn given by

$$(12) \qquad \begin{aligned} A &= \sigma_{cy} \big( \sigma_{by} \cosh(k_y h_1) \cosh(k_y d) + \sigma_{cy} \sinh(k_y h_1) \sinh(k_y d) \big) \\ B &= \sigma_{by} \cosh(k_y h_1) \sinh(k_y d) + \sigma_{cy} \sinh(k_y h_1) \cosh(k_y d) \\ C &= \sigma_{by} \sigma_{cy} \big( \sigma_{by} \sinh(k_y h_1) \cosh(k_y d) + \sigma_{cy} \cosh(k_y h_1) \sinh(k_y d) \big) \\ D &= \sigma_{by} \big( \sigma_{by} \sinh(k_y h_1) \sinh(k_y d) + \sigma_{cy} \cosh(k_y h_1) \cosh(k_y d) \big) \end{aligned}$$

Thus, from Eq (10) we can solve for $u_a$,

$$(13) \qquad u_a = \dfrac{e^{-k_{ay} y_0}}{2 k_{ay}} \cdot \left( \dfrac{A k_{ay} \sigma_{ay} - C k_y}{A k_{ay} \sigma_{ay} + C k_y} \right)$$

Finally, we can solve for $v_d$ in term of $u_a$,

$$(14) \qquad \begin{aligned} \dfrac{e^{-k_{ay} y_0}}{2} & \left( \dfrac{k_y D}{k_{ay}} - B \sigma_{ay} \right) + u_a \cdot \big( k_y D + k_{ay} \sigma_{ay} B \big) \\ &= v_d \cdot \big( \sigma_{by} \sigma_{cy} k_y e^{-k_y(h_1+d)} \big) \end{aligned}$$

that is,

$$(15) \qquad \begin{aligned} \dfrac{e^{-k_{ay} y_0}}{2 k_{ay}} & \left\{ \left( \dfrac{k_y D - k_{ay} \sigma_{ay} B}{k_y D + k_{ay} \sigma_{ay} B} \right) + \left( \dfrac{A k_{ay} \sigma_{ay} - C k_y}{A k_{ay} \sigma_{ay} + C k_y} \right) \right\} \\ &= v_d \cdot \left( \dfrac{\sigma_{by} \sigma_{cy} k_y e^{-k_y(h_1+d)}}{k_y D + k_{ay} \sigma_{ay} B} \right) \end{aligned}$$

which, using

$$AD - BC = \sigma_{by}^2 \sigma_{cy}^2$$

can finally be simplified to

(16) $$v_d\, e^{-k_y(h_1+d)} = e^{-k_{ay} y_0}\, \frac{\sigma_{ay}\, \sigma_{by}\, \sigma_{cy}}{A\, k_{ay}\, \sigma_{ay} + C\, k_y}$$

As pointed out by Farina and Rainoldi (1999) one can use this theoretical model to invert the filtering of the tissue and thus even resolve individual MUAPs from the sEMG data. These aspects will hopefully be developed further in the second part.